\documentclass[prd,preprint,tightenlines,floatfix,showpacs,preprintnumbers
,nofootinbib,superscriptaddress,eqsecnum]{revtex4}

 \usepackage[dvips,final]{graphicx}
  \usepackage{amssymb}
   \usepackage{amsmath}
    \usepackage{amsfonts}
     \usepackage{epsfig}
      \usepackage{bm}

\newcommand{\bPhi}{\mbox{\boldmath $\Phi$}}
\newcommand{\bkappa}{\mbox{\boldmath $\kappa$}}
\newcommand{\bl}{\mbox{\boldmath $l$}}

\newcommand{\bq}{\mbox{\boldmath $q$}}

\newcommand{\bk}{\mbox{\boldmath $k$}}
\newcommand{\bfb}{\mbox{\boldmath $b$}}
\newcommand{\bE}{\mbox{\boldmath $E$}}
\newcommand{\Jot}{{\cal{J}}}

\def\R  {{\cal R}}
\def\J {$J/\psi$ }

\def\cpc#1#2#3  {{Computer\ Phys.\ Comm.\ }  {\bf#1}, #2 (#3)}
\def\err#1#2#3  {{\it Erratum }              {\bf#1}, #2 (#3)}
\def\epjc#1#2#3 {{Eur. Phys. J. C }          {\bf#1}, #2 (#3)}
\def\dum#1#2#3  {{~}                         {\bf#1}, #2 (#3)}
\def\ib#1#2#3   {{\it ibid. }                {\bf#1}, #2 (#3)}
\def\jcp#1#2#3  {{J.\ Comput.\ Phys.\ }      {\bf#1}, #2 (#3)}
\def\jetpl#1#2#3 {{\rm JETP Lett.}           {\bf#1}, #2 (#3)}
\def\jhep#1#2#3 {{JHEP }                     {\bf#1}, #2 (#3)}
\def\ijmp#1#2#3 {{Int.\ J.\ Mod.\ Phys.\ }   {\bf#1}, #2 (#3)}
\def\jpg#1#2#3  {{J.\ Phys.\ G }             {\bf#1}, #2 (#3)}
\def\mpl#1#2#3  {{Mod.\ Phys.\ Lett.\ }      {\bf#1}, #2 (#3)}
\def\ncim#1#2#3 {{Nuovo Cimento }            {\bf#1}, #2 (#3)}
\def\np#1#2#3   {{Nucl.\ Phys.\ }            {\bf#1}, #2 (#3)}
\def\pan#1#2#3  {{Phys.\ At.\ Nuclei }       {\bf#1}, #2 (#3)}
\def\plb#1#2#3  {{Phys.\ Lett.\ B }          {\bf#1}, #2 (#3)}
\def\prep#1#2#3 {{Phys.\ Rep.\ }             {\bf#1}, #2 (#3)}
\def\prd#1#2#3  {{Phys.\ Rev.\ D }           {\bf#1}, #2 (#3)}
\def\prl#1#2#3  {{Phys.\ Rev.\ Lett.\ }      {\bf#1}, #2 (#3)}
\def\ptp#1#2#3  {{Prog.\ Theor.\ Phys.\ }    {\bf#1}, #2 (#3)}
\def\ps#1#2#3   {{Phpsica Scripta }          {\bf#1}, #2 (#3)}
\def\rmp#1#2#3  {{Rev.\ Mod.\ Phys.\ }       {\bf#1}, #2 (#3)}
\def\rpp#1#2#3  {{Rep.\ Prog.\ Phys.\ }      {\bf#1}, #2 (#3)}
\def\sa#1#2#3   {{Sci. Acta}                 {\bf#1}, #2 (#3)}
\def\sjnp#1#2#3 {{Sov.\ J.\ Nucl.\ Phys.\ }  {\bf#1}, #2 (#3)}
\def\spj#1#2#3  {{Sov.\ Phys.\ JETP }        {\bf#1}, #2 (#3)}
\def\spjl#1#2#3 {{Sov.\ JETP Lett.\ }        {\bf#1}, #2 (#3)}
\def\spu#1#2#3  {{Sov.\ Phys.-Usp.\ }        {\bf#1}, #2 (#3)}
\def\yaf#1#2#3  {{Yad.\ Fiz.\ }              {\bf#1}, #2 (#3)}
\def\zp#1#2#3   {{Zeit.\ Phys.\ }            {\bf#1}, #2 (#3)}
\def\zpc#1#2#3  {{Z.\ Phys.\ C }             {\bf#1}, #2 (#3)}

\begin{document}

\title{The $\gamma \gamma \to J/\psi J/\psi$ reaction and
the $J/\psi J/\psi$ pair production \\ 
in exclusive ultraperipheral ultrarelativistic heavy ion collisions}

\author{Sergey Baranov}
\email{baranov@sci.lebedev.ru}
\affiliation{P.N. Lebedev Institute of Physics, 53 Lenin Avenue, Moscow
  119991, Russia}

\author{Anna Cisek}
\email{anna.cisek@ifj.edu.pl} 
\affiliation{Institute of Nuclear
Physics PAN, PL-31-342 Cracow, Poland}

\author{Mariola K{\l}usek-Gawenda}
\email{mariola.klusek@ifj.edu.pl}
\affiliation{Institute of Nuclear
Physics PAN, PL-31-342 Cracow, Poland}

\author{Wolfgang Sch\"afer}
\email{wolfgang.schafer@ifj.edu.pl} 
\affiliation{Institute of Nuclear
Physics PAN, PL-31-342 Cracow, Poland}

\author{Antoni Szczurek}
\email{antoni.szczurek@ifj.edu.pl} 
\affiliation{Institute of Nuclear
Physics PAN, PL-31-342 Cracow,
Poland}
\affiliation{
University of Rzesz\'ow, PL-35-959 Rzesz\'ow, Poland}

\date{\today}

\begin{abstract}
We calculate the cross section for the $\gamma \gamma \to J/\psi J/\psi$ process.
Two mechanisms are considered: box (two-loop) diagrams of the order of
$O(\alpha_{em}^2 \alpha_s^2)$ and two-gluon exchange of the order of 
$O(\alpha_{em}^2 \alpha_s^4)$. The first mechanism is calculated in the 
heavy-quark non-relativistic approximation while the second case we
also include the effects of quantum motion of quarks in the bound state. 
The box contribution dominates at energies close to the threshold 
($W <$ 15 GeV) while the two-gluon mechanism takes over at $W >$ 15 GeV.
Including the bound-state wave function effects for the two-gluon exchange 
mechanism gives a cross section 0.1 - 0.4 pb, substantially smaller 
than that in the non-relativistic limit (0.4 - 1.6 pb). 
We also find a strong infrared 
sensitivity which manifests itself in a rather strong dependence on the
mass for the $t$-channel gluons.
The elementary cross section is then used in the Equivalent Photon
Approximation (EPA) in the impact parameter space 
to calculate the cross section for 
$^{208}Pb+^{208}Pb \to ^{208}Pb + J/\psi J/\psi + ^{208}Pb$
reaction. Distributions in rapidity of the $J/\psi J/\psi$ pair and 
invariant mass of the pair are shown. 
\end{abstract}


\maketitle

\section{Introduction}

The two-photon collisions is a natural place to study vector-meson pair
production. Production of light vector mesons is rather of nonperturbative
nature and therefore subjected to more phenomenological studies.
Heavy vector meson production ($J/\psi$ or $\Upsilon$) is particularly 
interesting as here the pQCD degrees of freedom can be applied and 
hopefully reliable predictions can be obtained.
It was realized quite early that in the high-energy limit the two-gluon
exchange is the dominant reaction mechanism of two heavy vector meson 
production.
The relevant amplitude was calculated first in the heavy-quark
non-relativistic approximation in \cite{GPS88}, and first estimates
of the cross section were presented there.
The impact factors of \cite{GPS88} were used next to estimate BFKL effects
e.g. in \cite{KM98,GS2005}. Some other estimates based on parametrizing
dipole-dipole interaction were presented
in \cite{GM2007}.
Perhaps surprisingly, there is actually a substantial spread in the
predictions shown in those works, they
do not seem consistent one with each other and differ by almost 
two orders of magnitude. For a number of reasons,
it seems that these previous evaluations are not
very realistic in the intermediate energy range relevant to 
studies of the $\gamma \gamma$ sub-process in nucleus-nucleus collisions.

Firstly, the process considered in the mentioned papers is not 
the lowest order of strong coupling constant.
Formally lower-order processes of the box type were studied
in \cite{Qiao}. In these calculation heavy-quark approximation
was used in order to calculate the two-loop amplitudes.
At high energies, these mechanisms correspond to the quark-antiquark 
exchanges in the crossed channels and will die out 
with increasing $\gamma \gamma$ energies.
The numerical predictions of \cite{Qiao} were however 
shown only at high (linear collider) energies
($\sqrt{s} >$ 400 GeV) where this mechanism is definitely not the
dominant one. 

Secondly, previous calculations of the two-gluon exchange mechanism
were restricted to the extreme non-relativistic limit, and neglect 
the motion of heavy quarks in the bound state.
Here, we shall calculate the two-gluon exchange contribution 
using for the first time for this process 
relativistic non-forward impact factors.
Furthermore, by introducing a gluon mass, 
we investigate the infrared sensitivity of the amplitude,
which turns out to be rather strong.

Therefore in the present work we wish to discuss both mechanisms 
simultaneously and discuss and identify the region of their dominance.

Finally we shall present realistic predictions for heavy ion collisions
where the fluxes of quasi-real photons are very large.
Here one may be able to study such processes in the near future.

\section{Formalism}

\begin{figure}[!h]
\includegraphics[width=2.5cm]{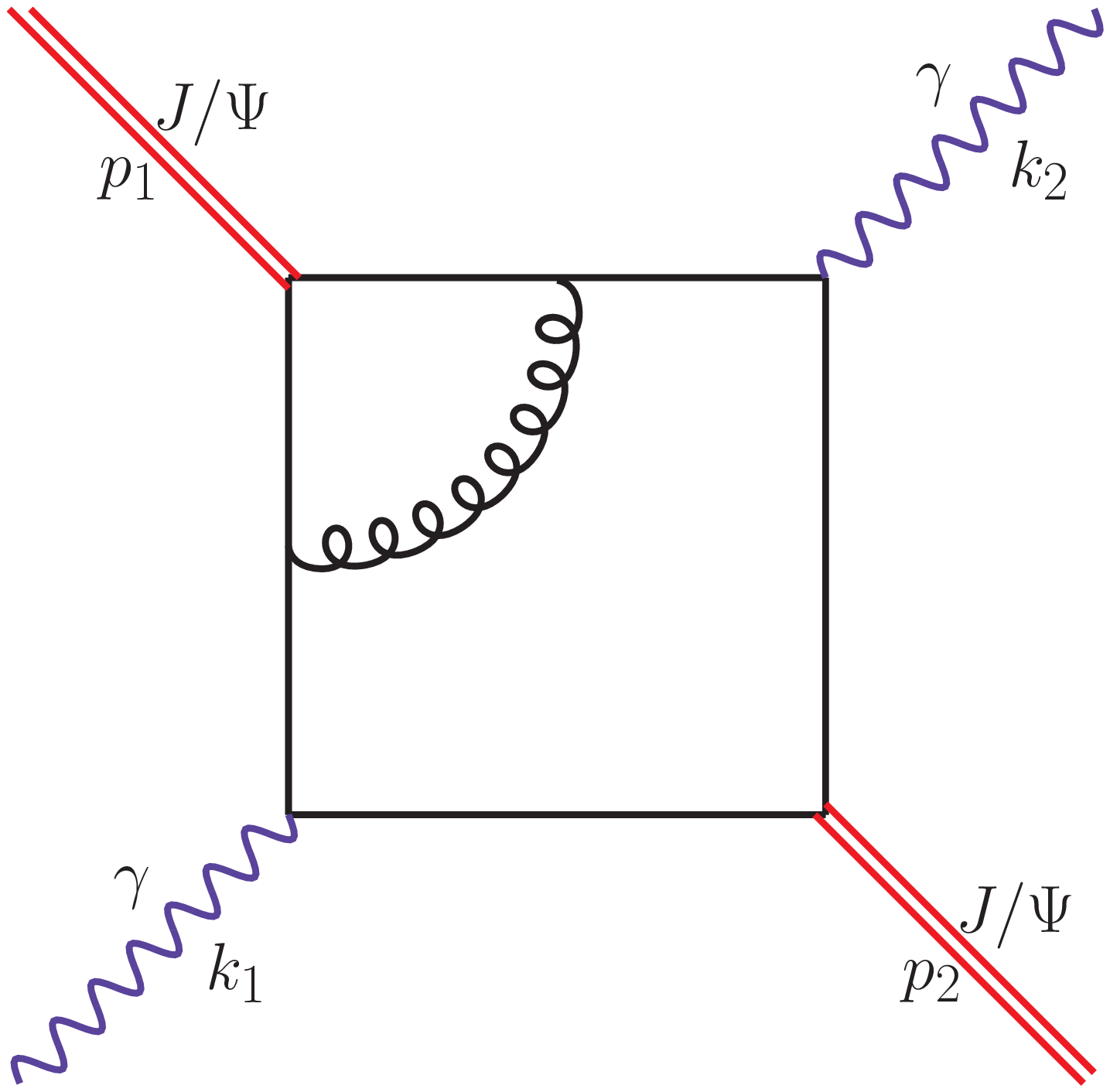}
\includegraphics[width=2.5cm]{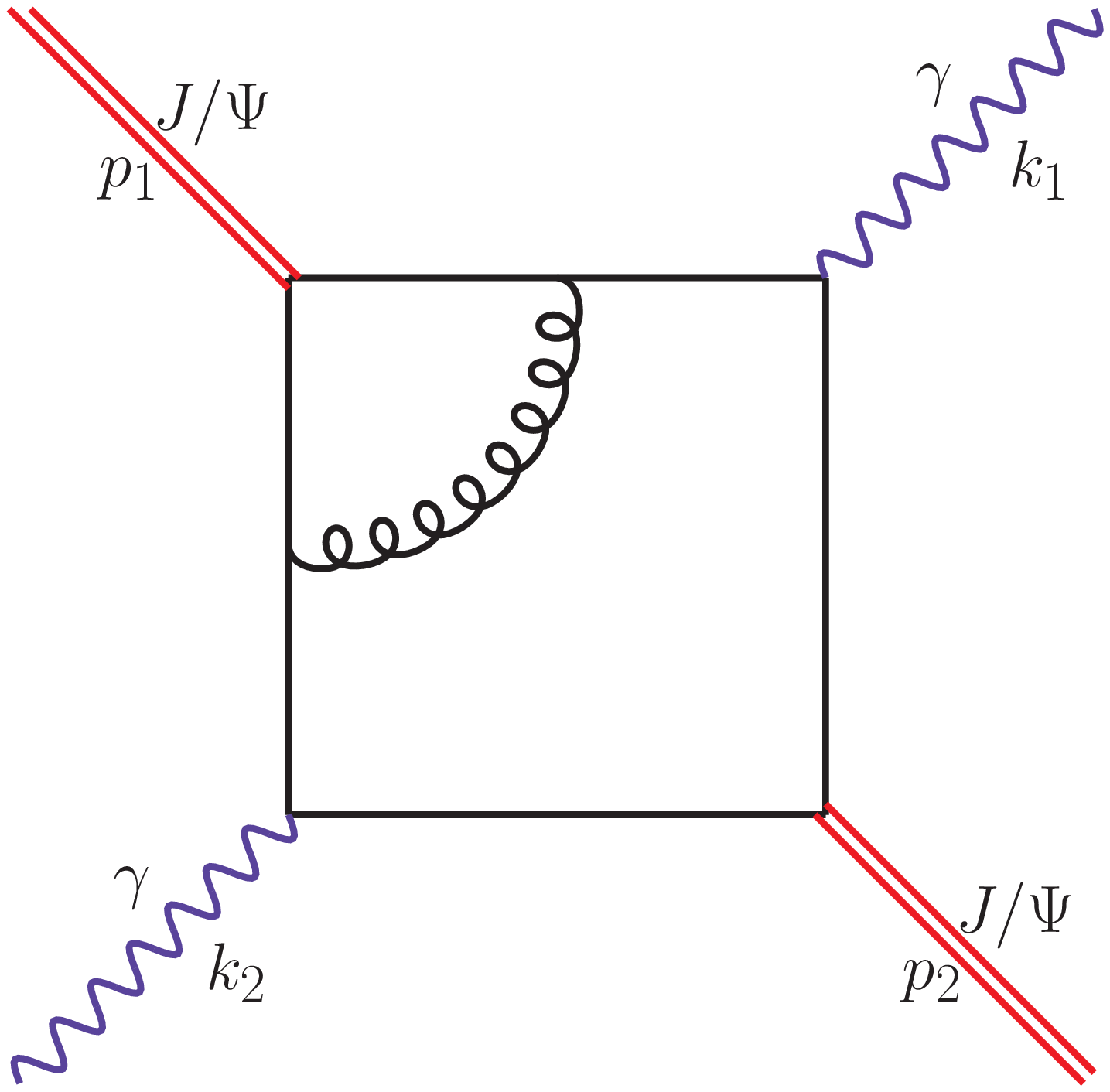}
\includegraphics[width=2.5cm]{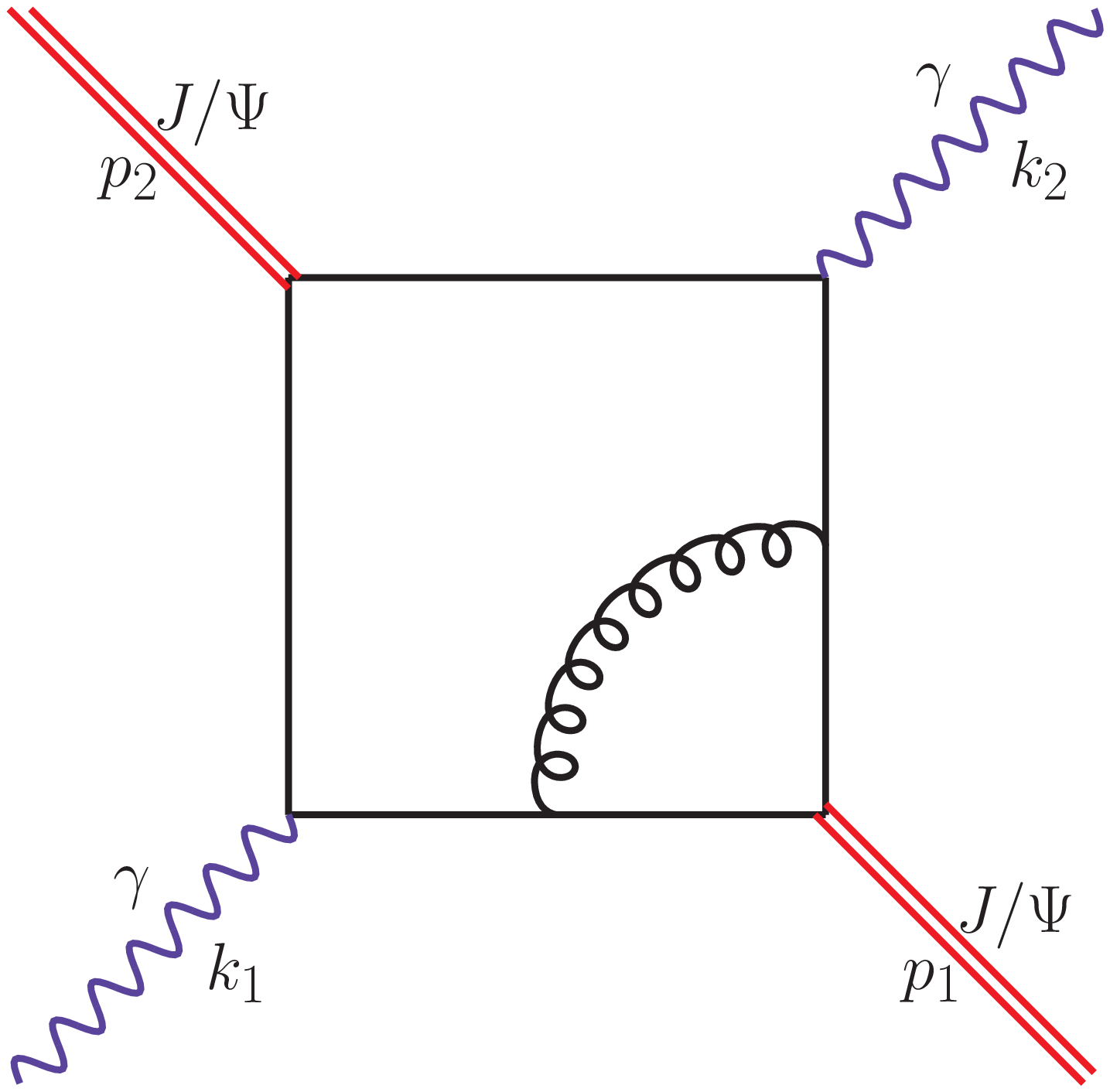}
\includegraphics[width=2.5cm]{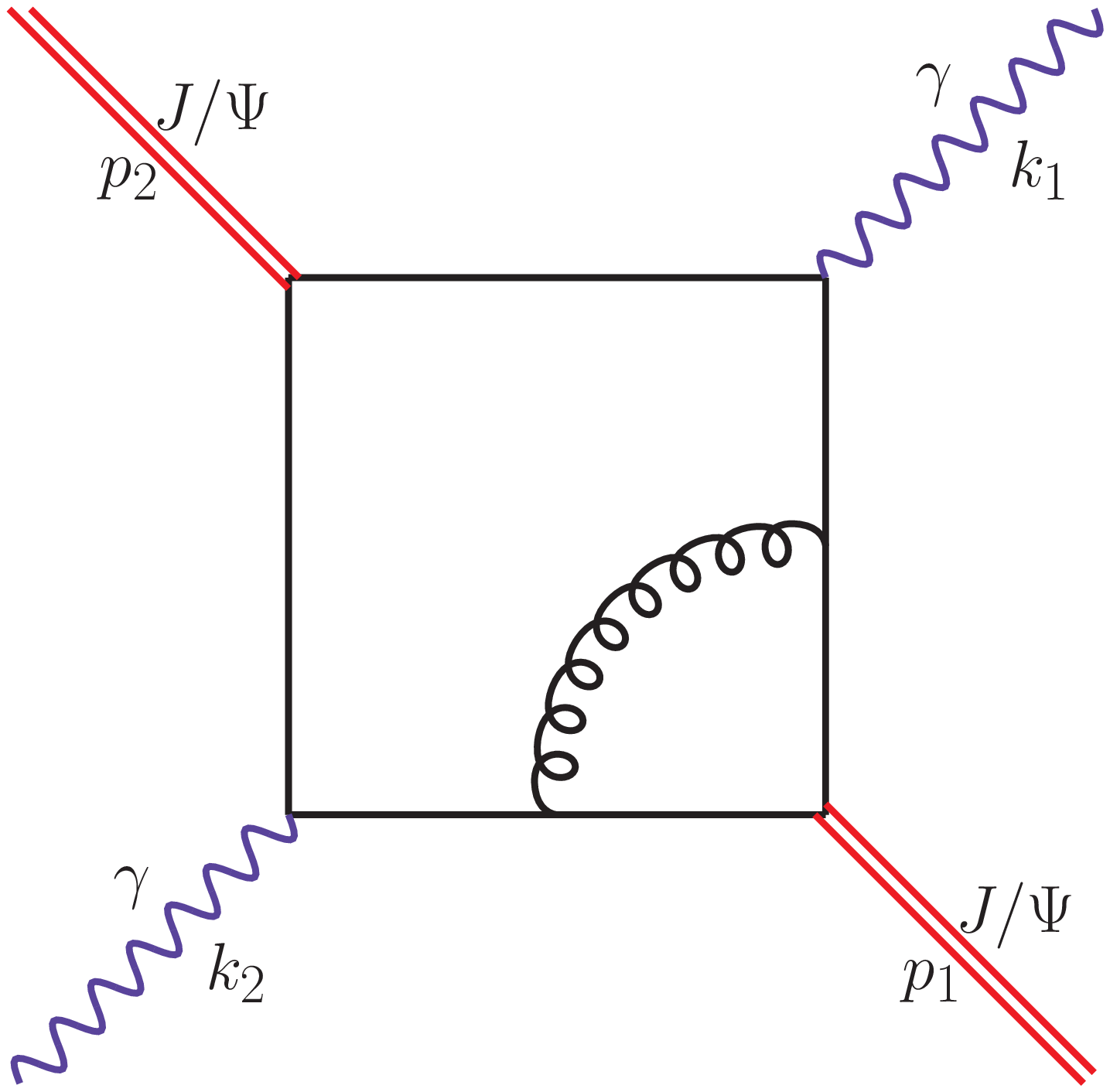}\\
\includegraphics[width=2.5cm]{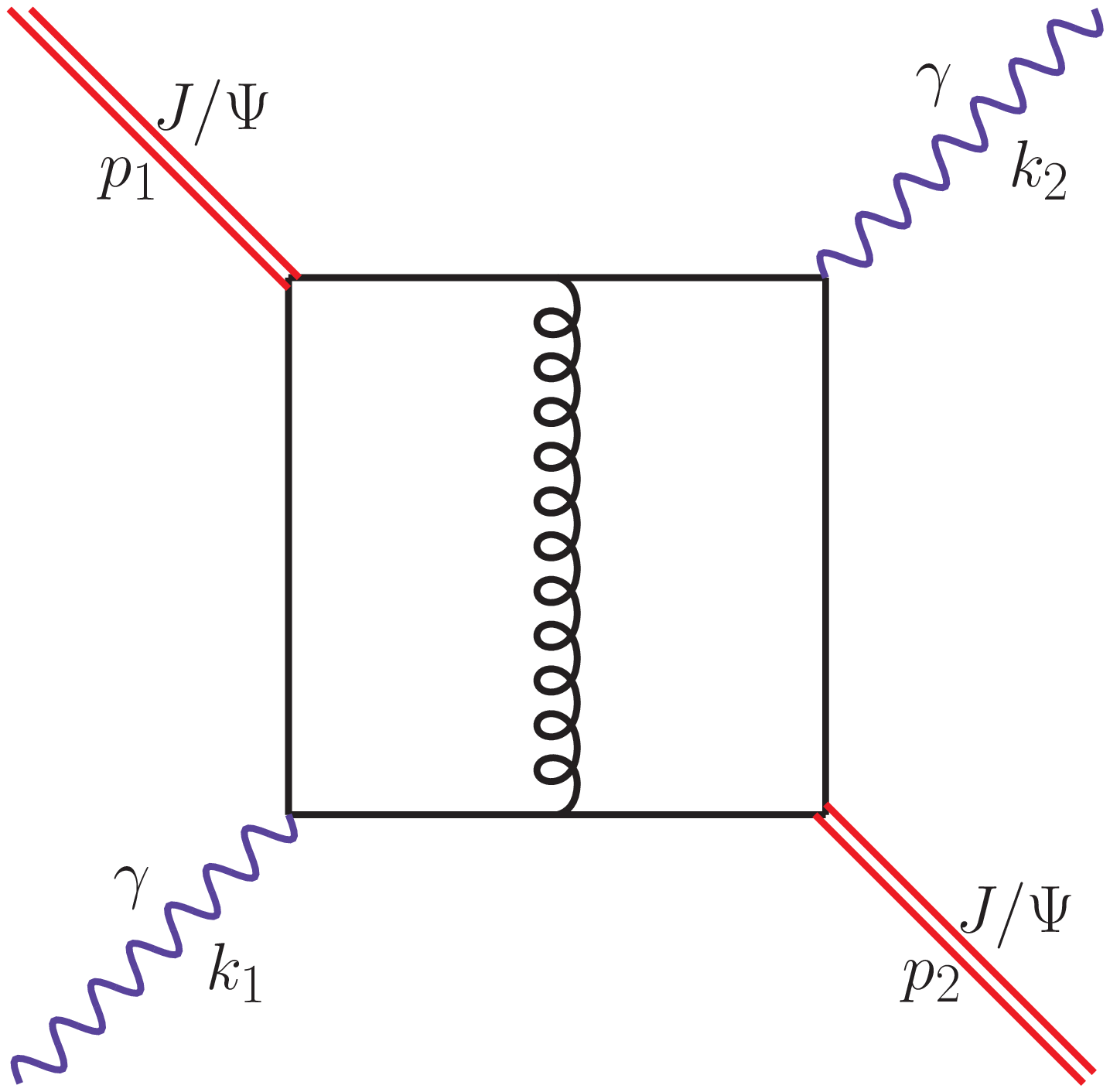}
\includegraphics[width=2.5cm]{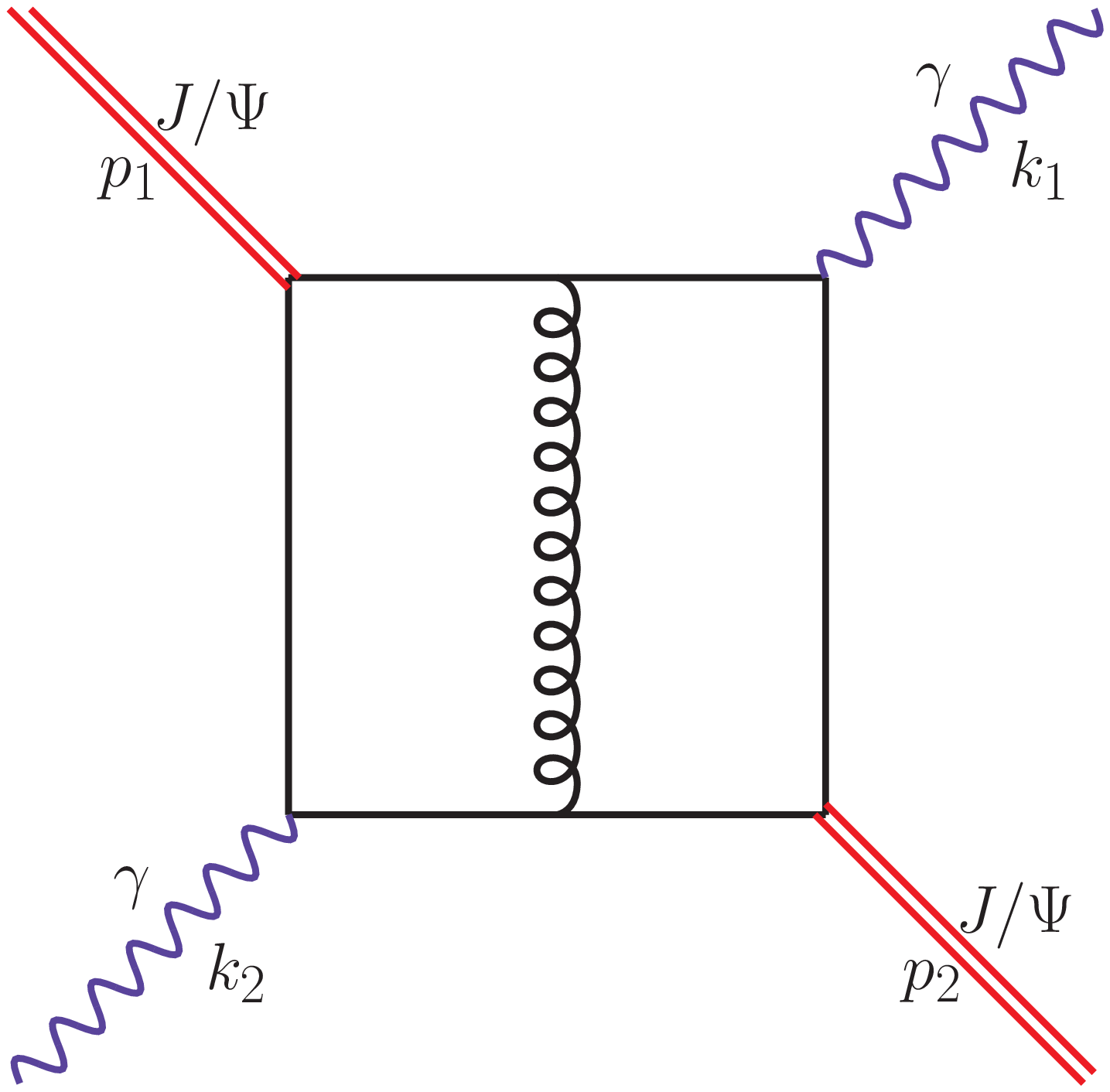}
\includegraphics[width=2.5cm]{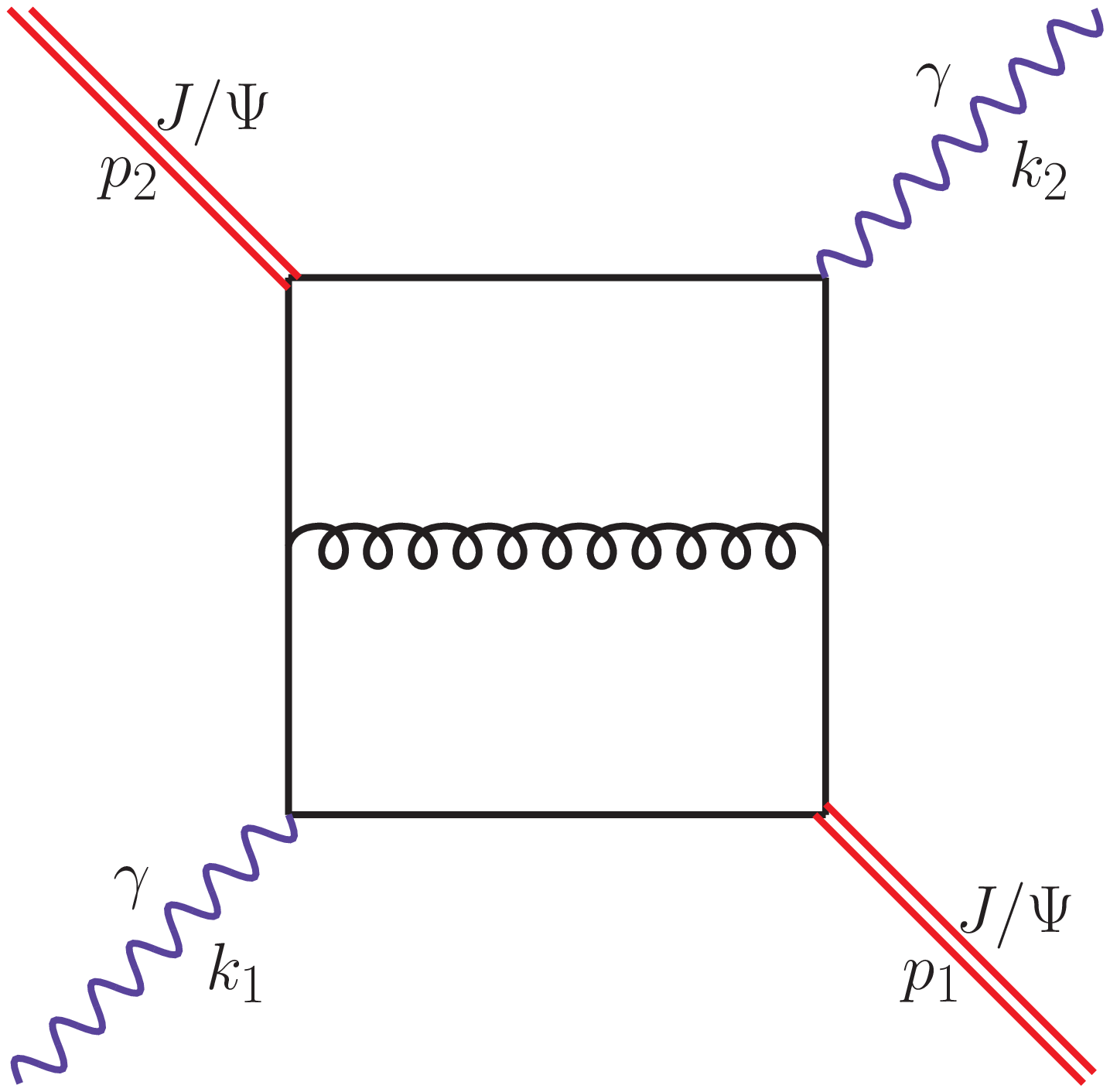}
\includegraphics[width=2.5cm]{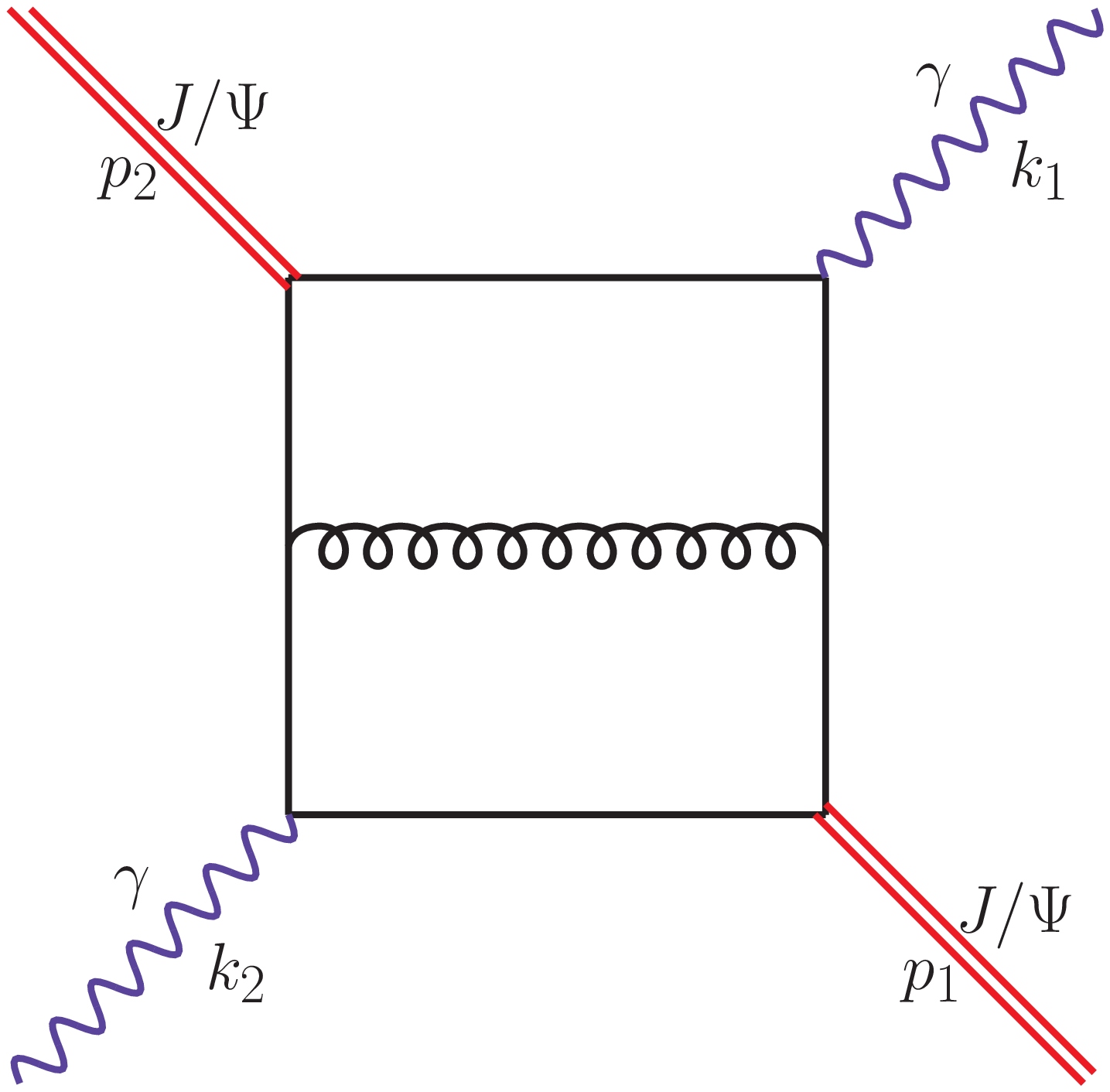}\\
\includegraphics[width=2.5cm]{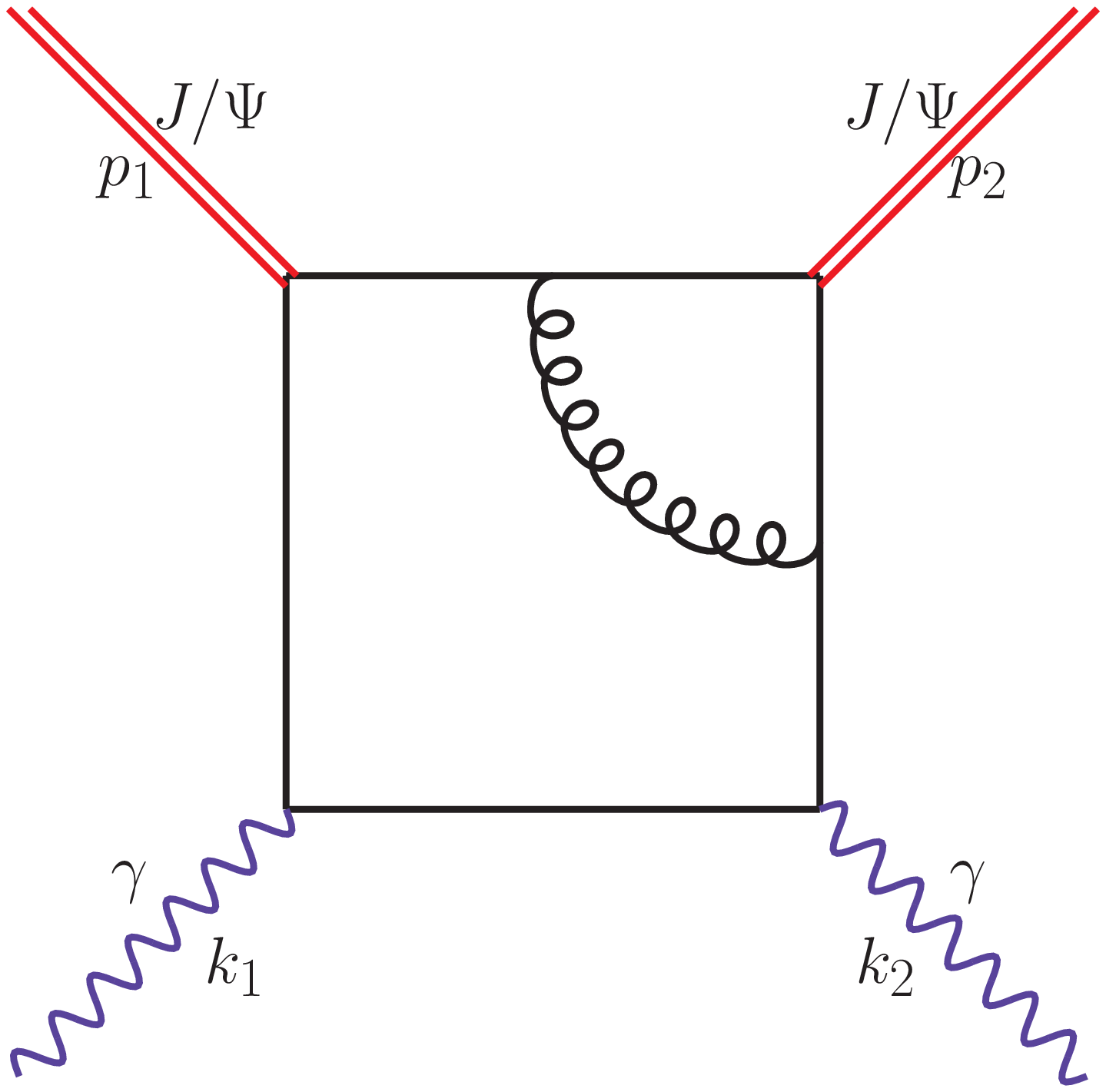}
\includegraphics[width=2.5cm]{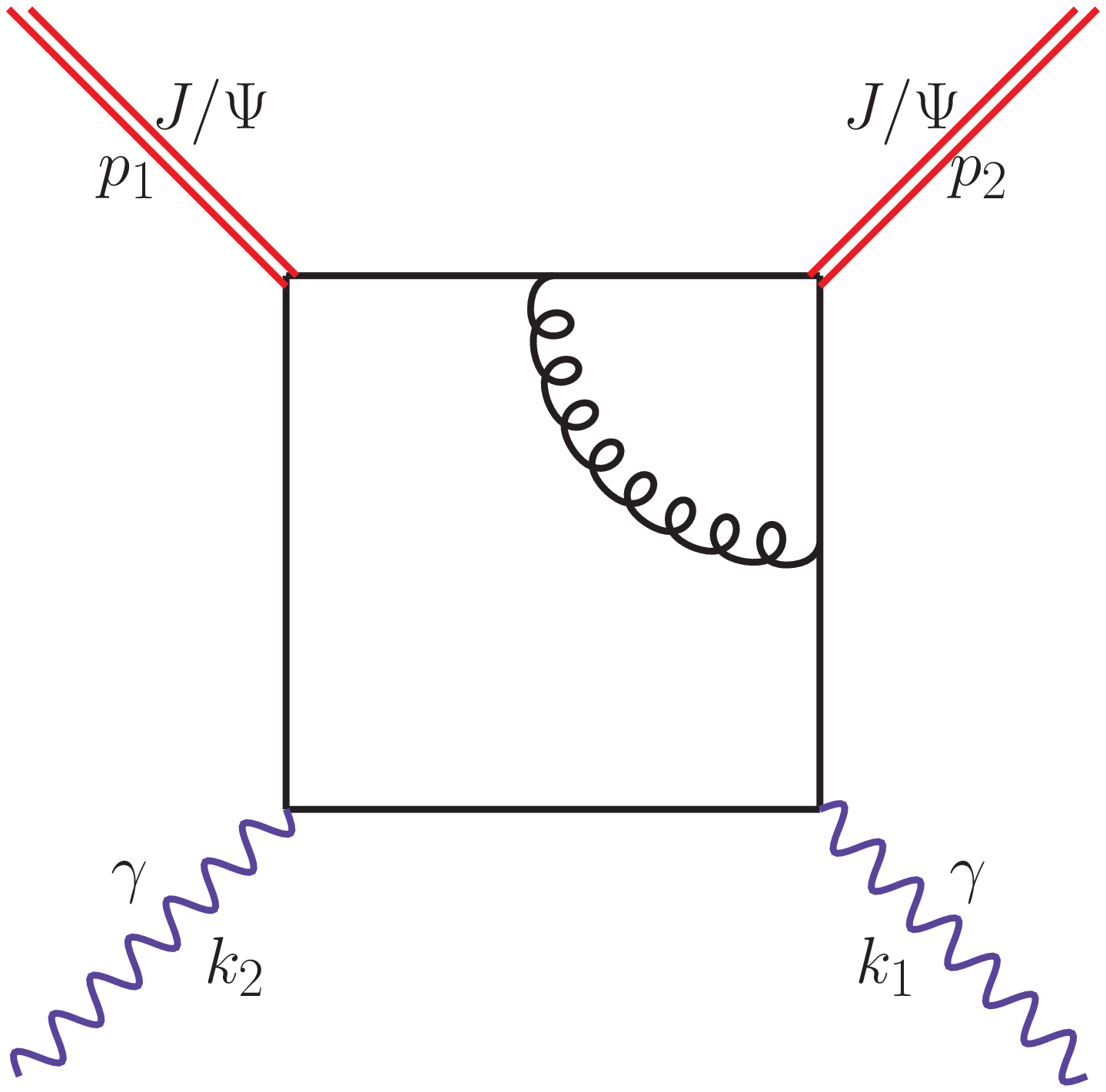}
\includegraphics[width=2.5cm]{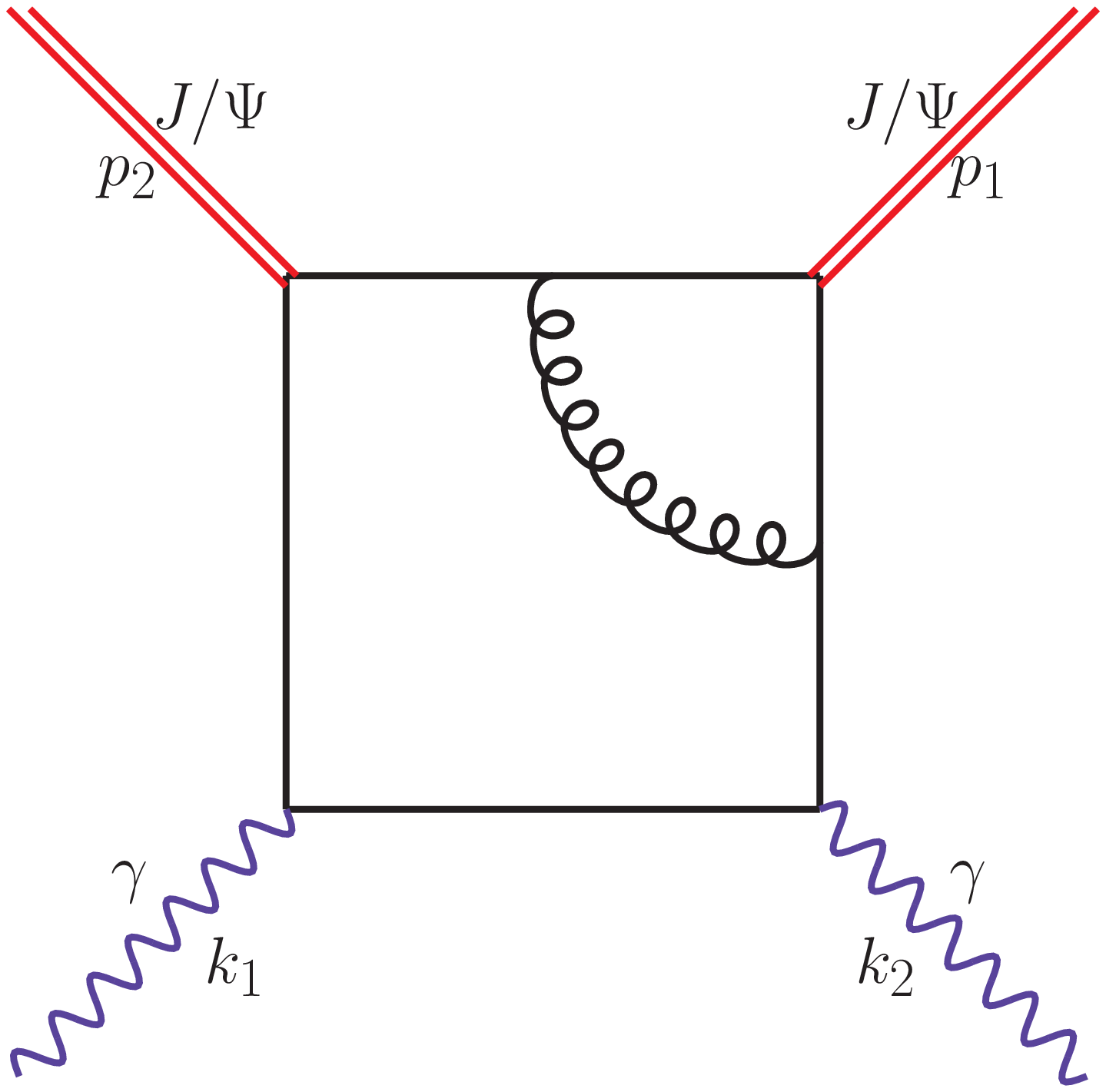}
\includegraphics[width=2.5cm]{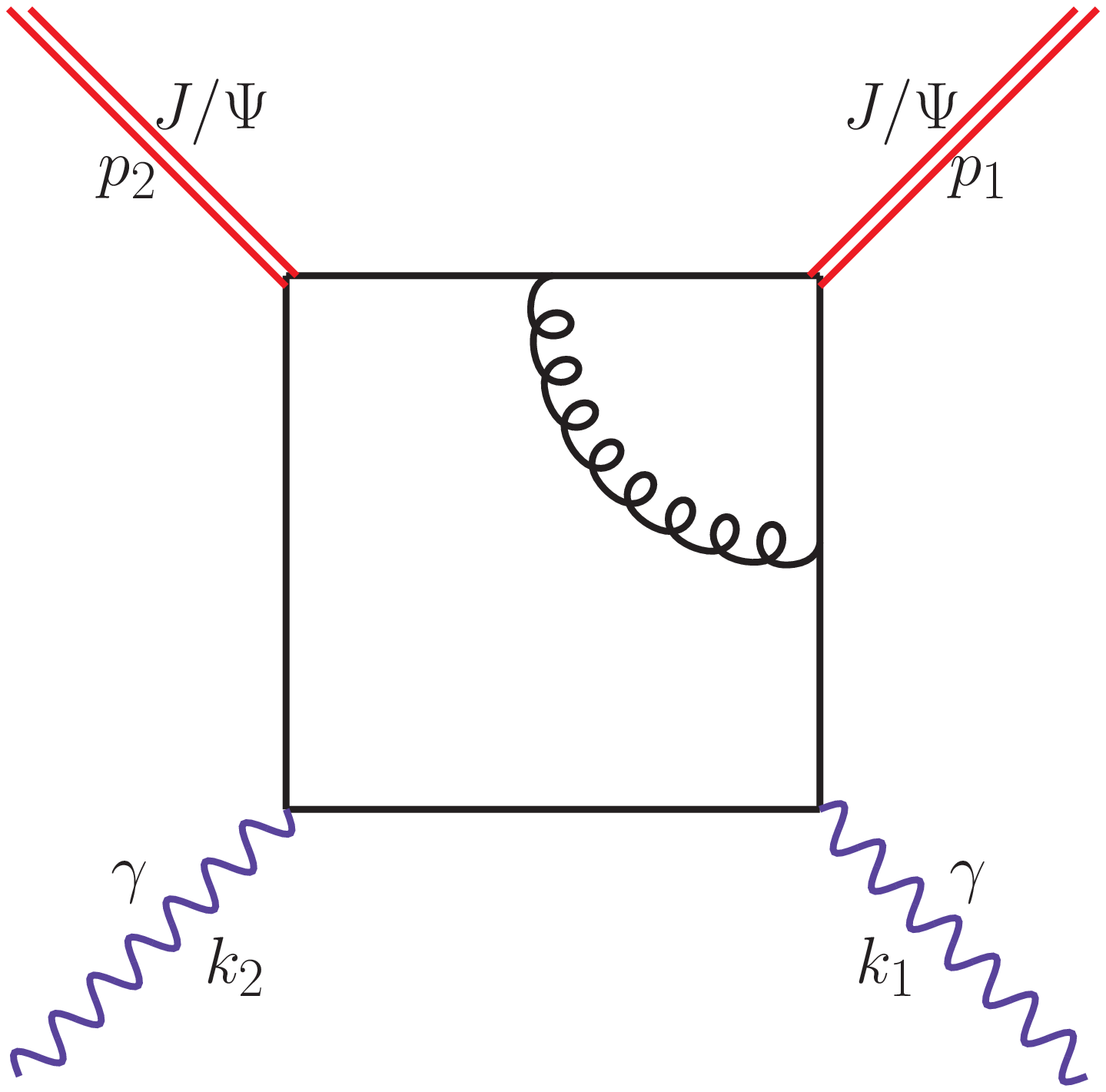}\\
\includegraphics[width=2.5cm]{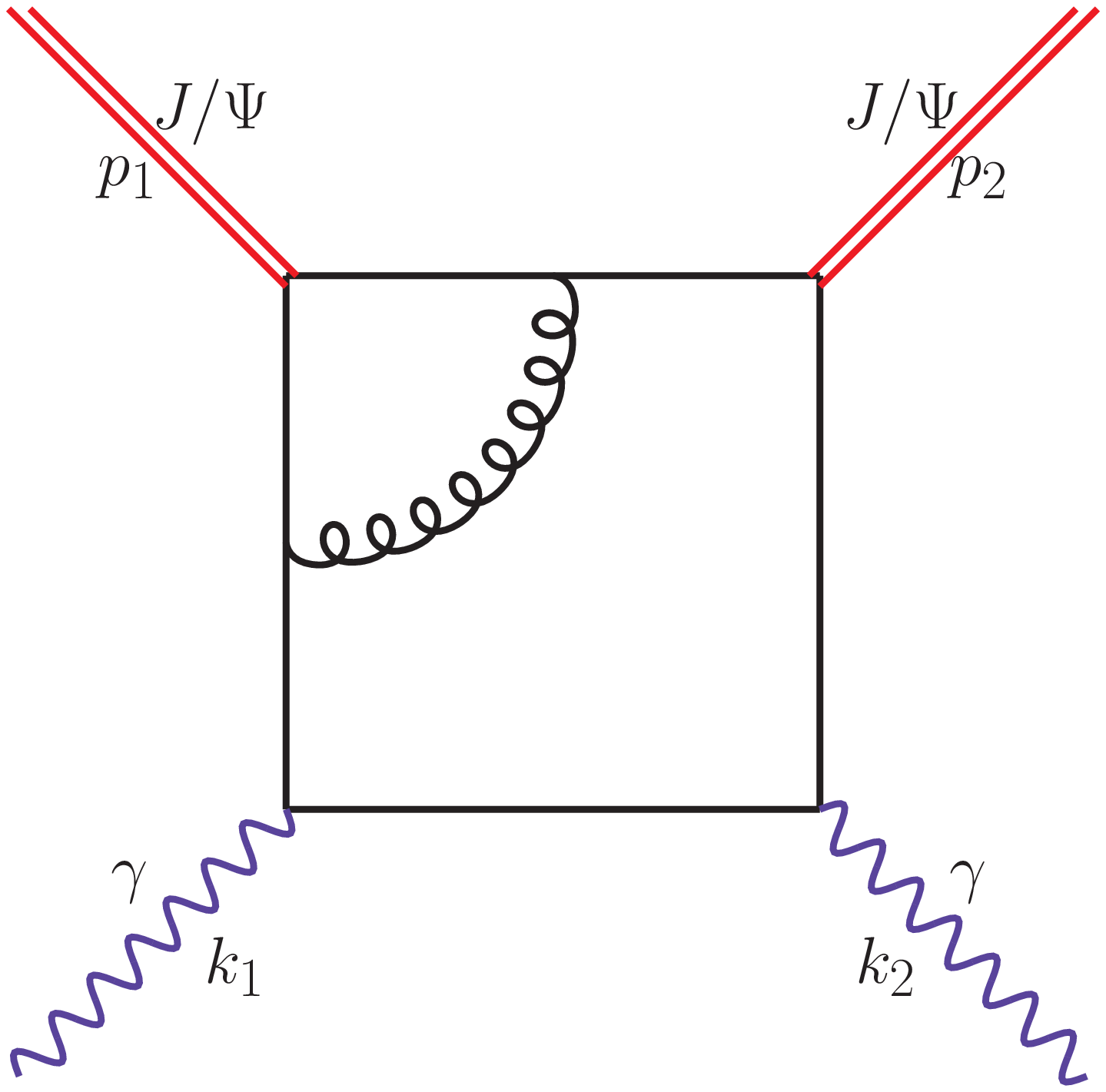}
\includegraphics[width=2.5cm]{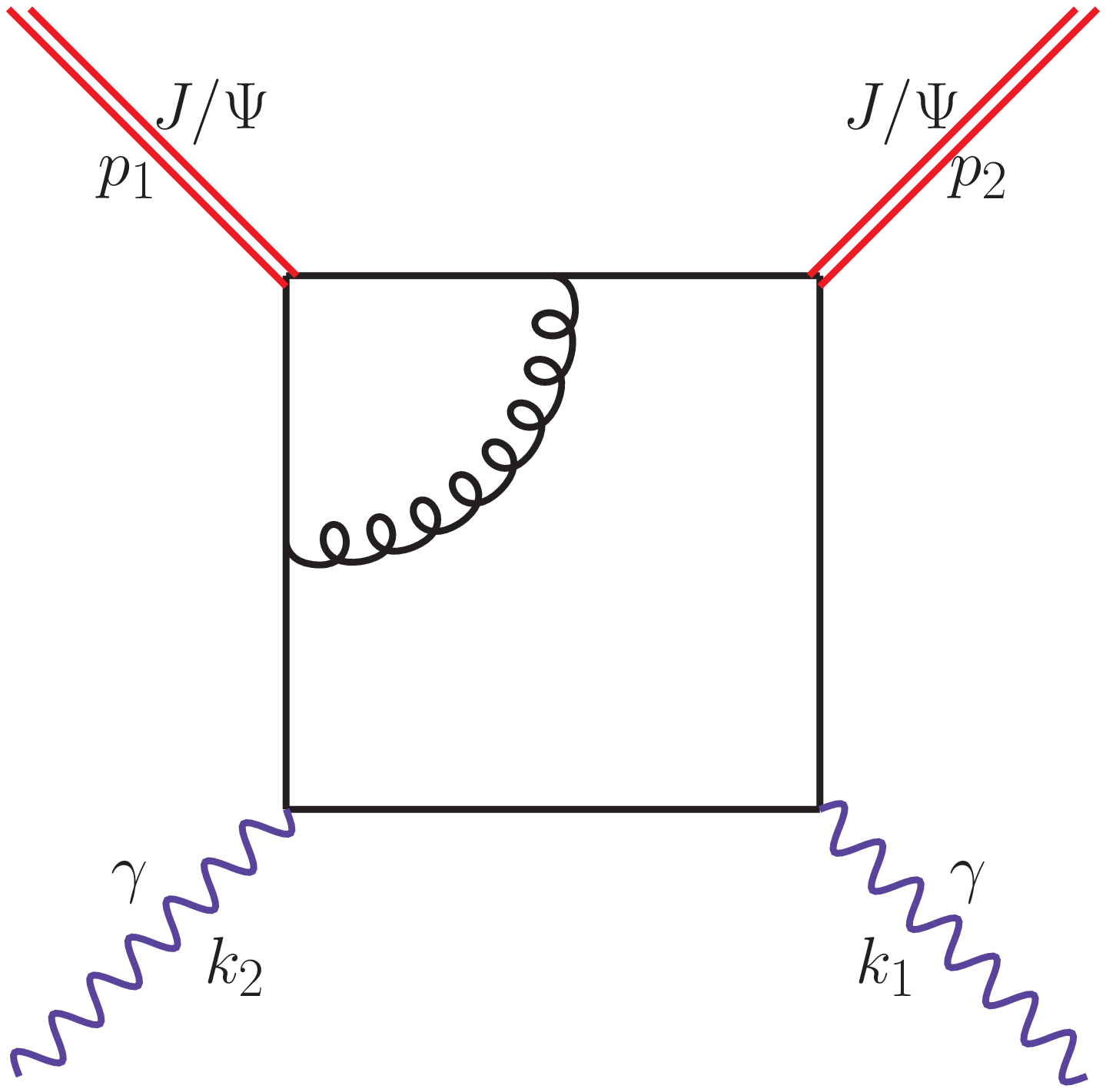}
\includegraphics[width=2.5cm]{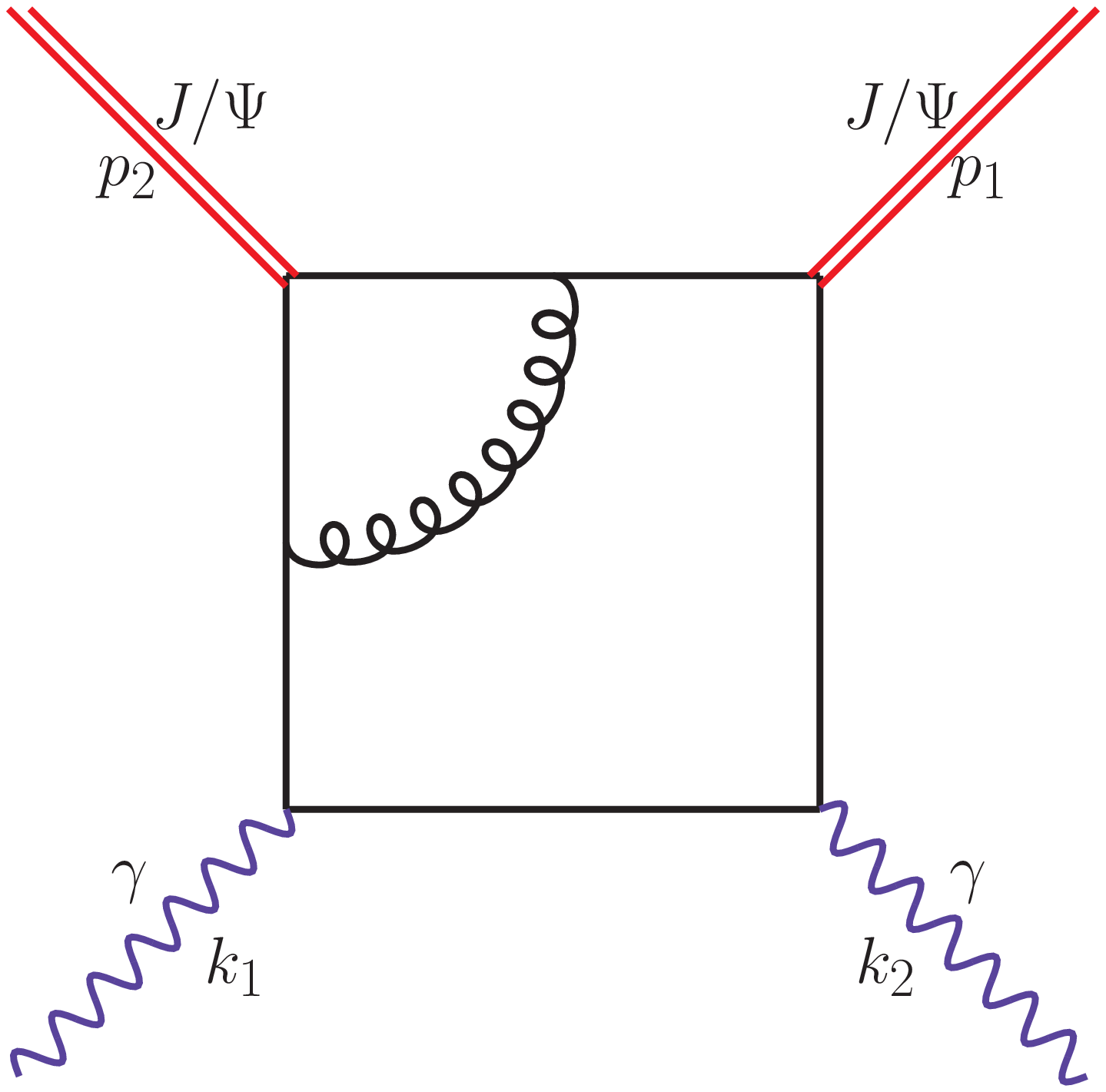}
\includegraphics[width=2.5cm]{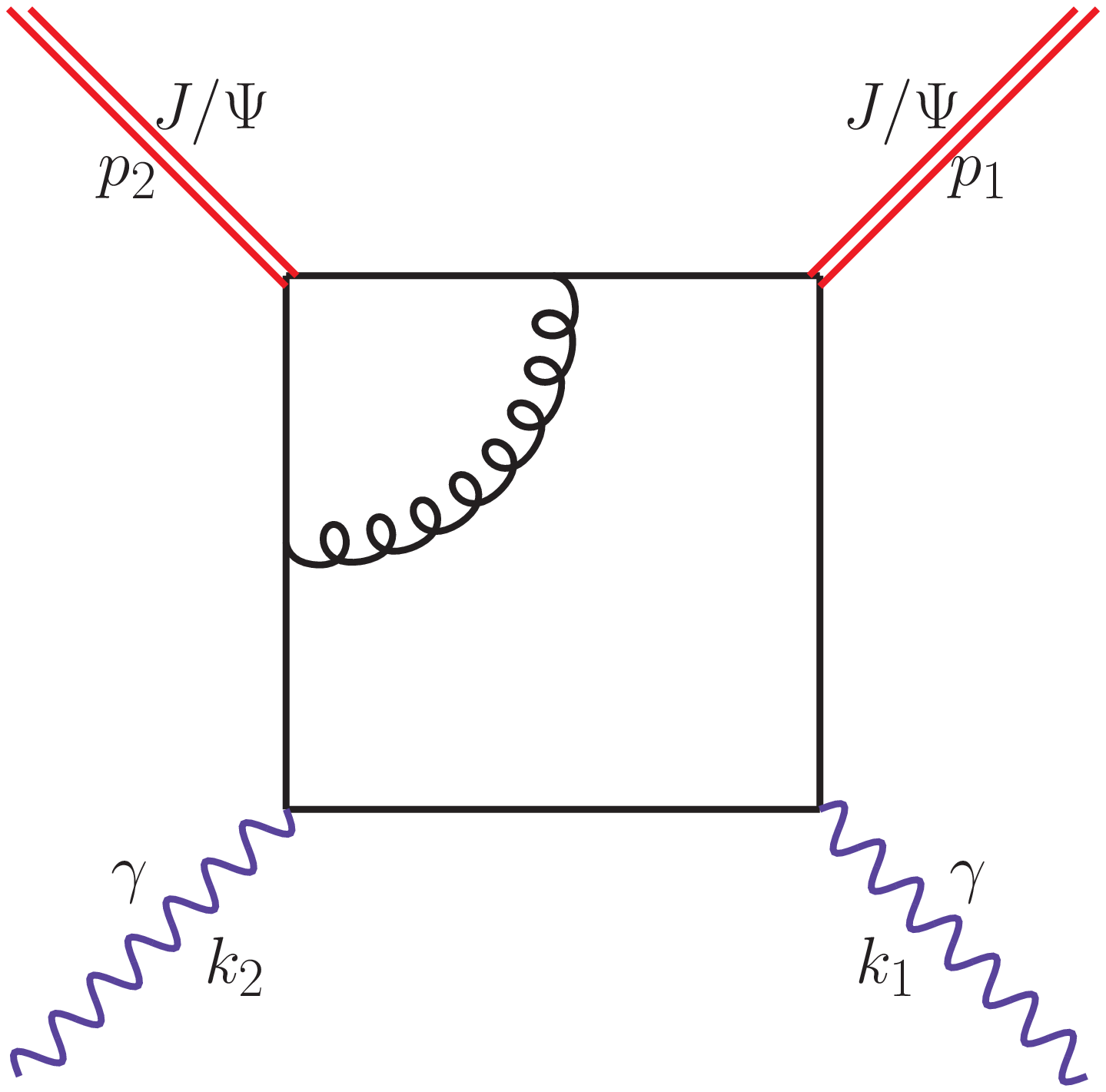}\\
\includegraphics[width=2.5cm]{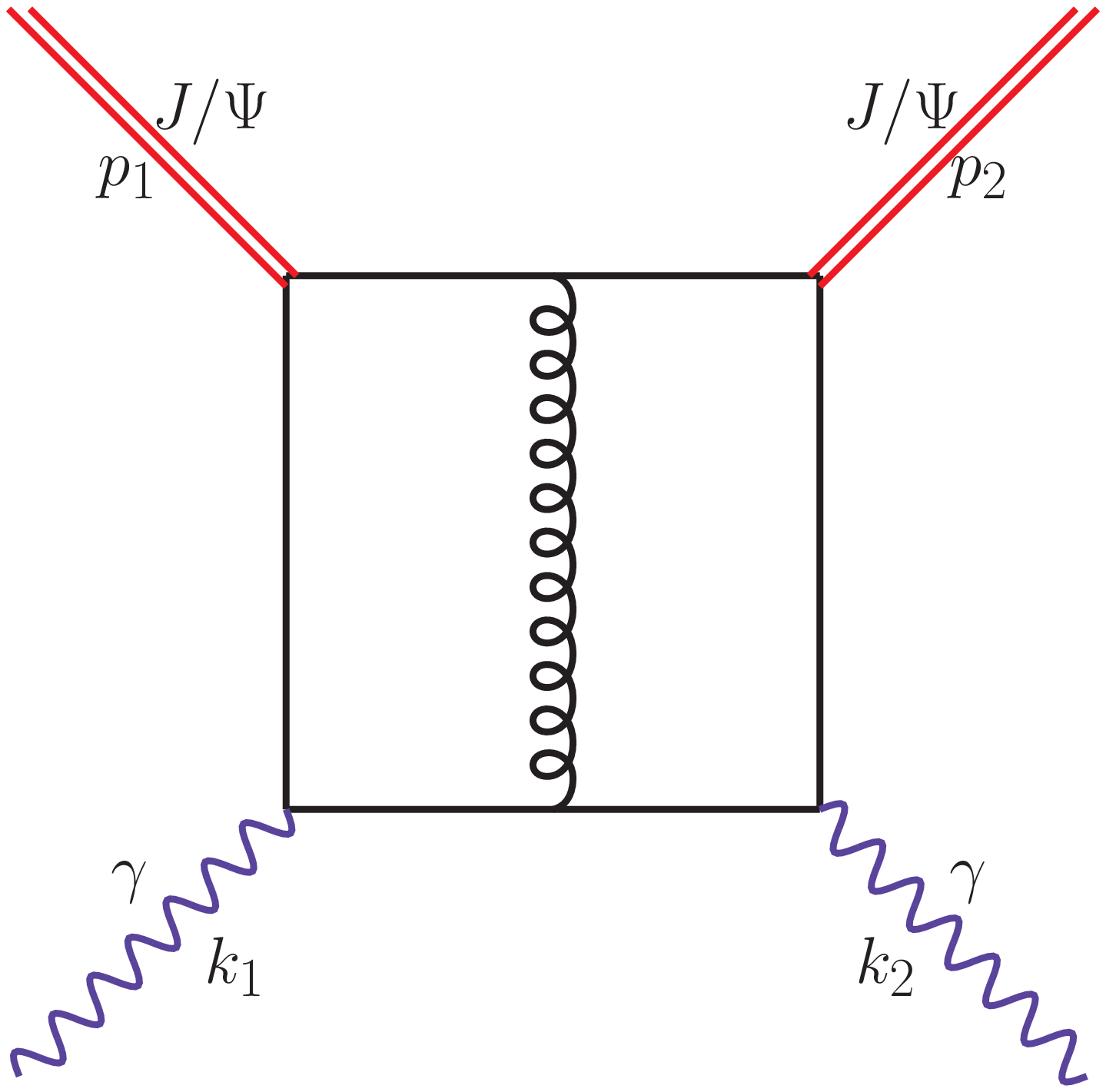}
\includegraphics[width=2.5cm]{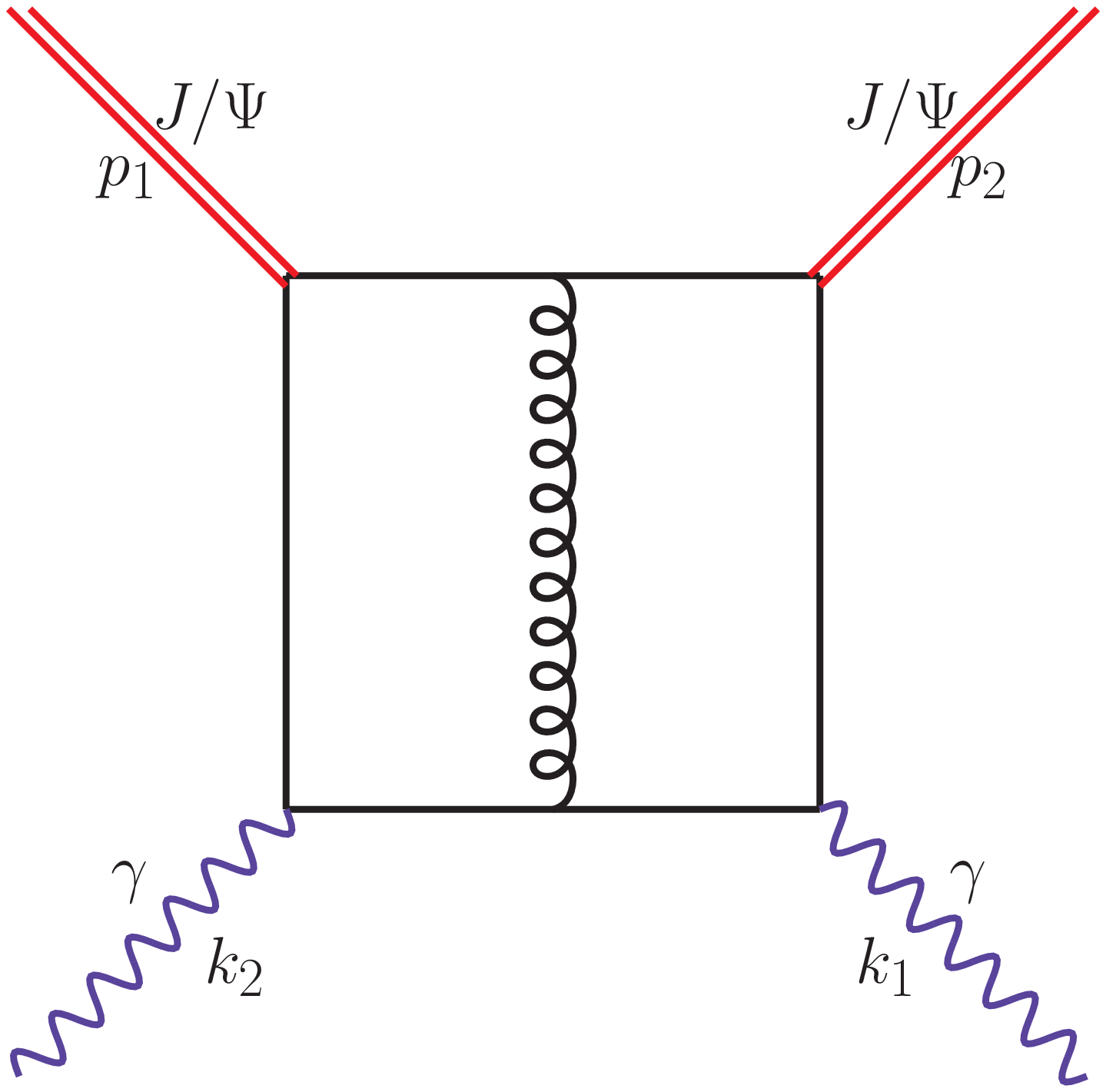}
\includegraphics[width=2.5cm]{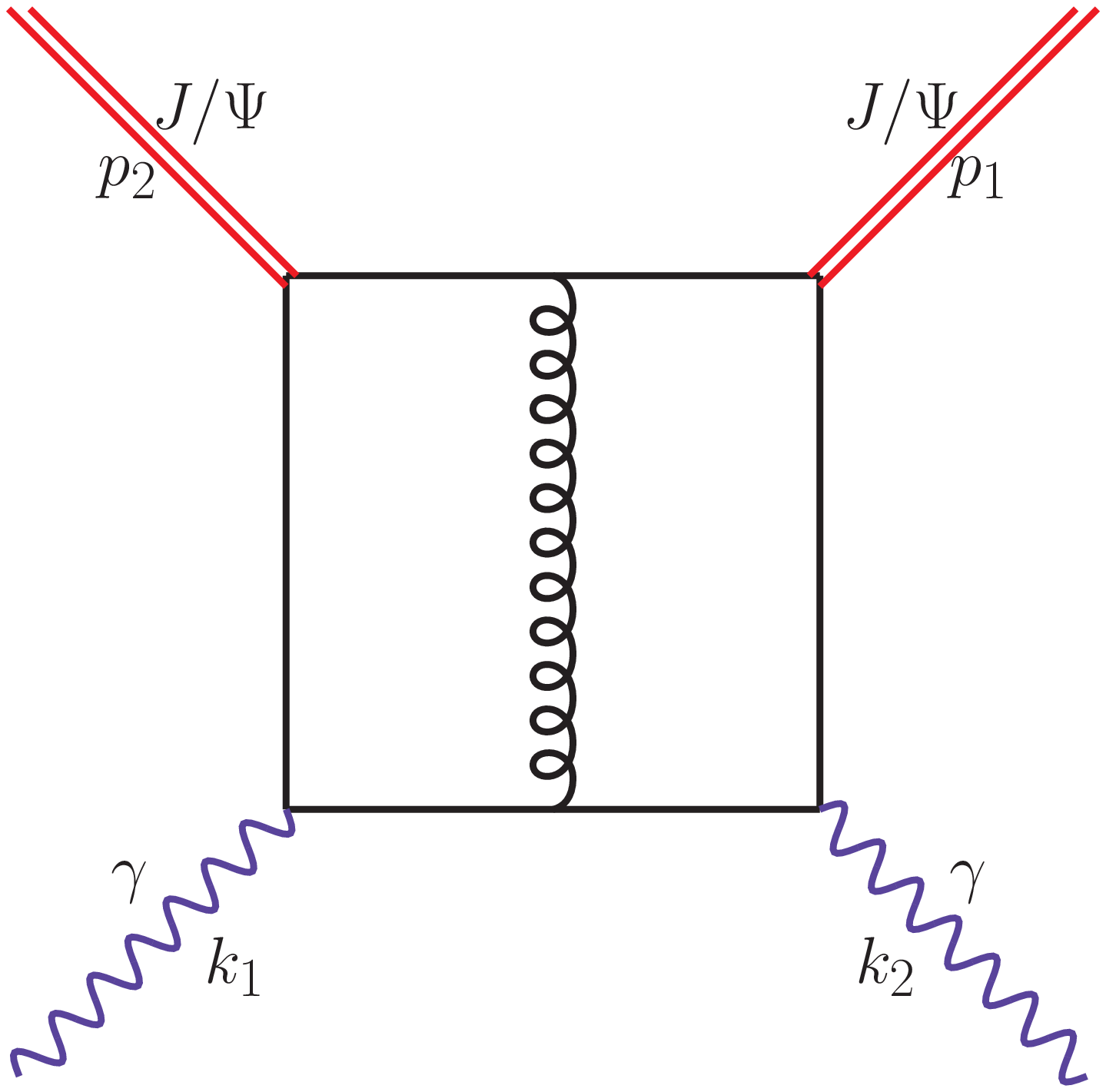}
\includegraphics[width=2.5cm]{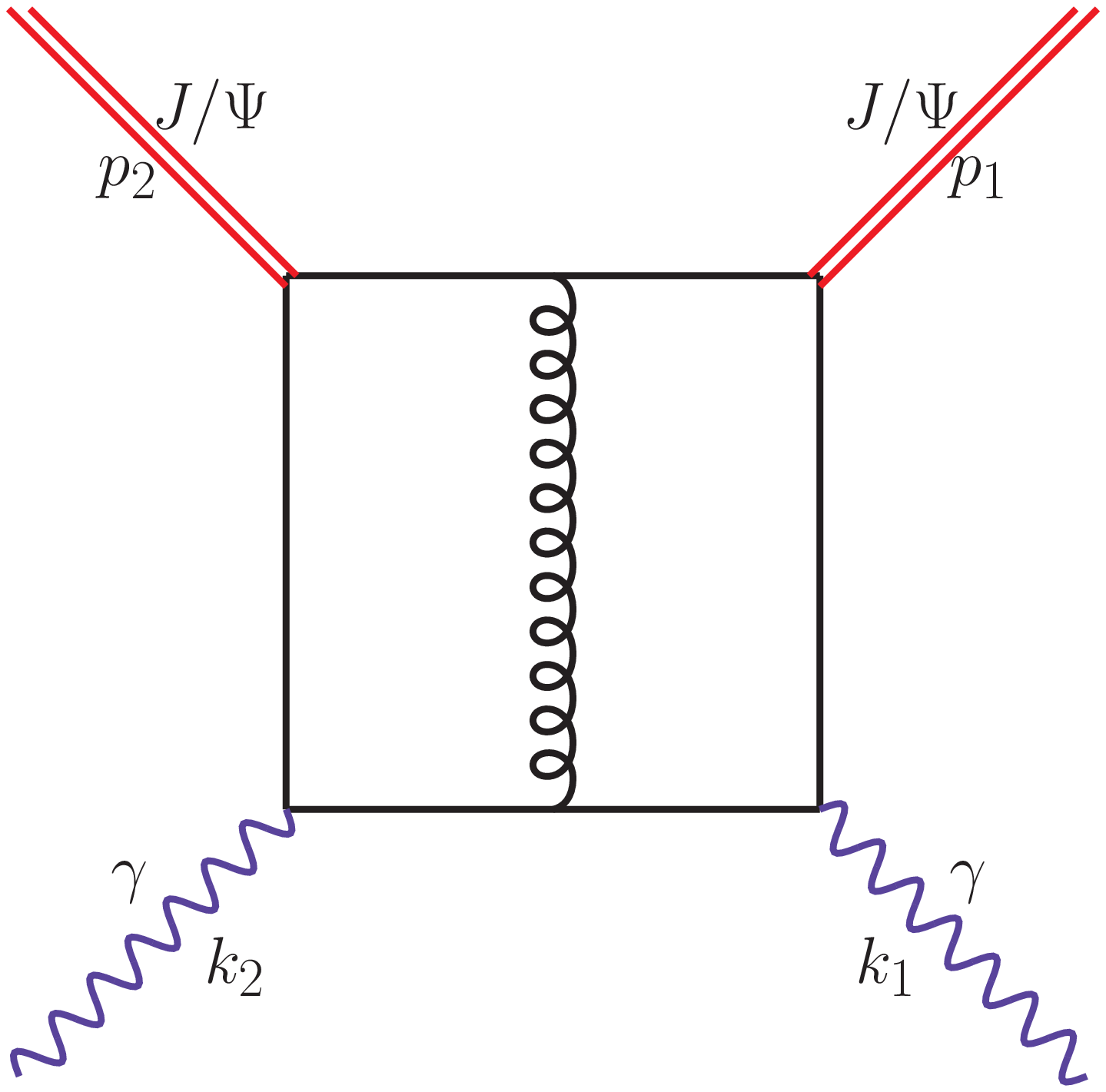}
   \caption{
\small The ``box''  diagrams for $\gamma \gamma \to J/\psi J/\psi$
included in the present paper in the heavy-quark approximation.
}
\label{fig:box_diagrams}
\end{figure}

\begin{figure}[!h]
\includegraphics[width=6cm]{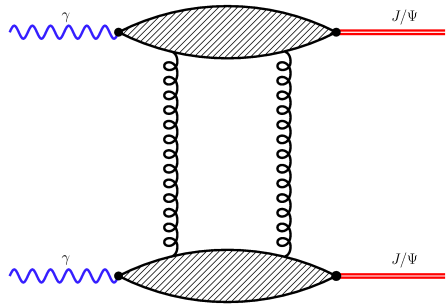}
   \caption{
\small Graphical representation of the two-gluon exchange mechanism for 
$\gamma \gamma \to J/\psi J/\psi$.
}
 \label{fig:two-gluon_exchange}
\end{figure}

\begin{figure}[!h]
\begin{center}
\includegraphics[width=4cm]{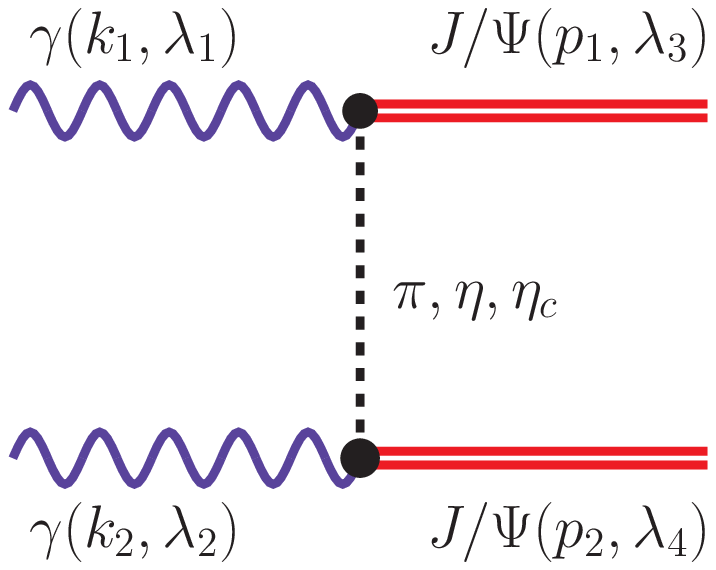}
\includegraphics[width=4cm]{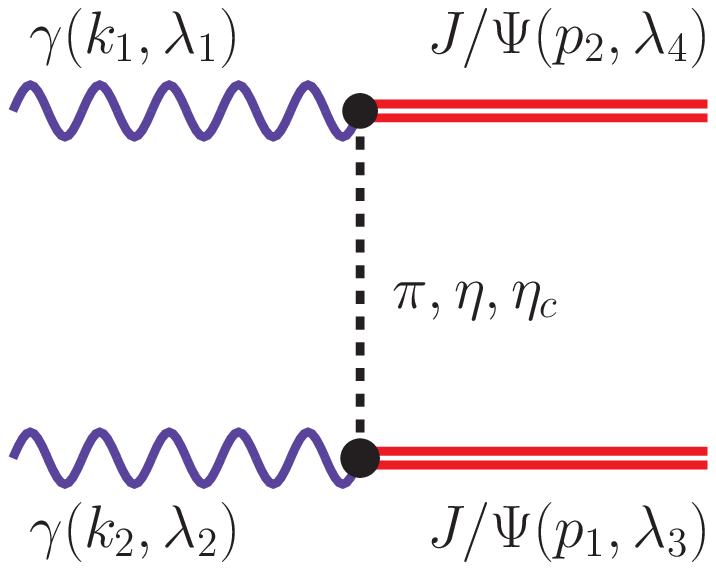}
\end{center}
   \caption{
\small Graphical representation of the meson exchange diagrams for 
$\gamma \gamma \to J/\psi J/\psi$.
}
 \label{fig:meson_exchange}
\end{figure}

In the present approach we include processes shown in Figs.
\ref{fig:box_diagrams}, \ref{fig:two-gluon_exchange}, \ref{fig:meson_exchange}.
Now we shall discuss each of the mechanisms one by one.

\subsection{Box diagrams}

At the leading order, ${\cal O}(\alpha^2\alpha_s^2)$,
the sub-process $\gamma+\gamma\to J/\psi+J/\psi$
is represented by the 20 "box" diagrams shown in Fig.\ref{fig:box_diagrams}
The calculations are straightforward and follow the standard QCD rules
and non-relativistic bound state formalism described in detail in
Refs. \cite{NRQCD}. As usual, the production amplitudes
contain spin and color projection operators that guarantee the proper
quantum numbers of the final state mesons.
The formation of the $c\bar{c}$ bound states is determined by the
\J wave function at the origin of coordinate space, $|\psi(0)|^2$;
its value is known from the leptonic decay width:
 $|\psi(0)|^2=|\R(0)|^2/(4\pi)=0.08$ GeV$^3$ \cite{PDG}.
Only the color-singlet channels are taken into consideration in the
present study.

Our calculation is identical to that of
Ref. \cite{Qiao}, with the only exception that we consider individual
helicity amplitudes rather than the entire spin-averaged matrix element.
The polarization vectors of the initial photons and outgoing mesons are
presented in our approach as explicit 4-vectors.
For a particle moving along the $z$-axis, $p^\mu=(0,\;0,\;|p|,\;E)$,
the helicity eigenstate vectors have the form
$$\epsilon^\mu(\pm 1)=(\pm 1,\;i,\;0,\;0)/\sqrt{2},
\quad \epsilon^\mu(0)= (0,\;0,\;E,\;|\vec{p}|)/m.$$
The evaluation of the Feynman diagrams has been done using the algebraic
manipulation system {\sc form} \cite{FORM}.
We have checked that after performing numerical summation over all
possible polarization states we arrive at the same result as the one
given by Eq.(4) in Ref. \cite{Qiao}.

In practical calculation we take leading-order formula for 
$\alpha_s$ which is evaluated at $\mu_r^2 = 4 m_c^2$. 

\subsection{Two-gluon exchange}

At sufficiently high energies, the box diagram contributions,
which contain the fermion-antifermion exchange in the crossed 
channels die out, and the cross section will be dominated by
diffractive mechanisms. The Feynman diagrams for 
the diffractive $\gamma \gamma \to VV$ amplitude 
is depicted in Fig.\ref{fig:two-gluon_exchange}. Although it 
is formally of higher order in $\alpha_S$ than the box mechanism, the 
crossed channel gluon exchanges do not die out with energy.

\subsubsection{Relativistic approach}

\begin{figure}[!h]
\begin{center}
\includegraphics[width=5cm]{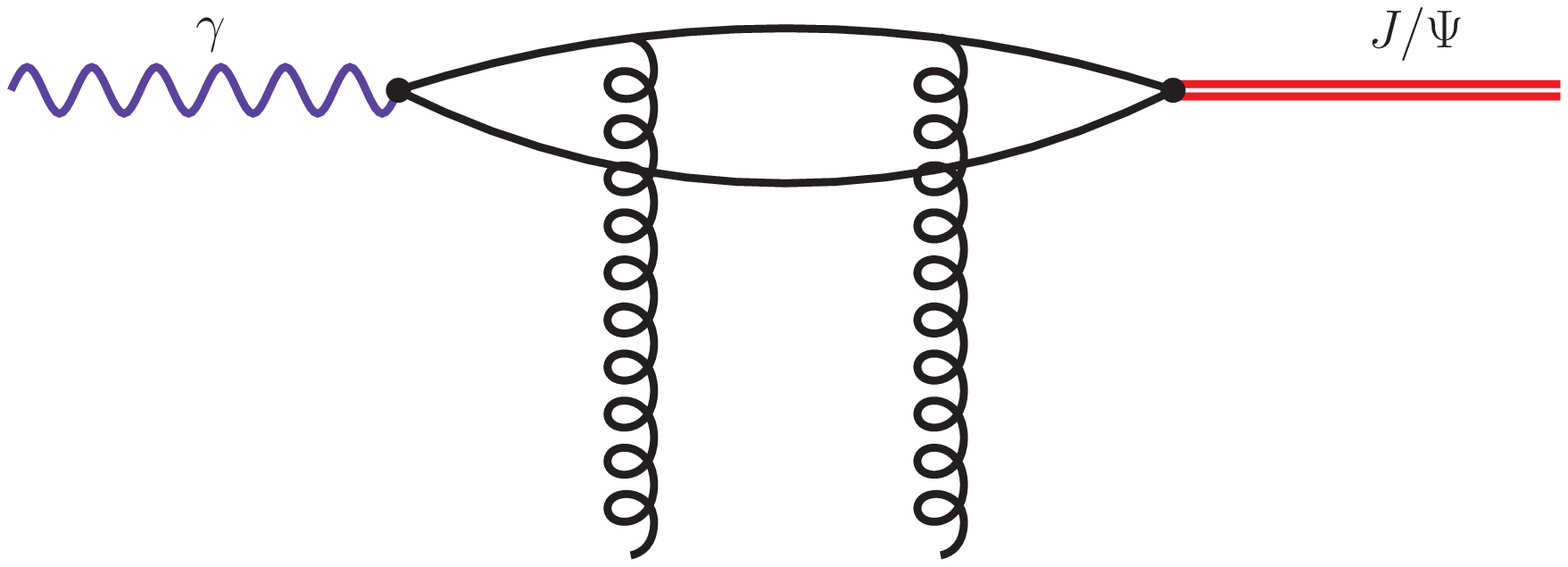}
\includegraphics[width=5cm]{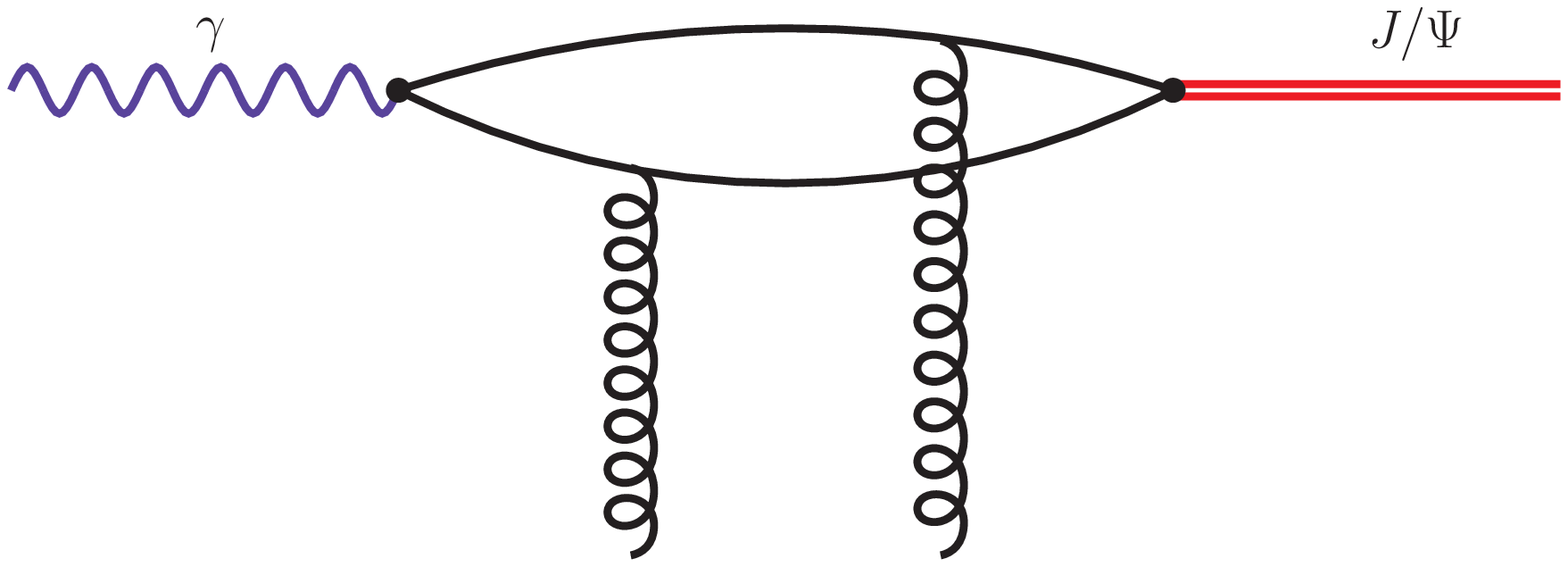}\\
\includegraphics[width=5cm]{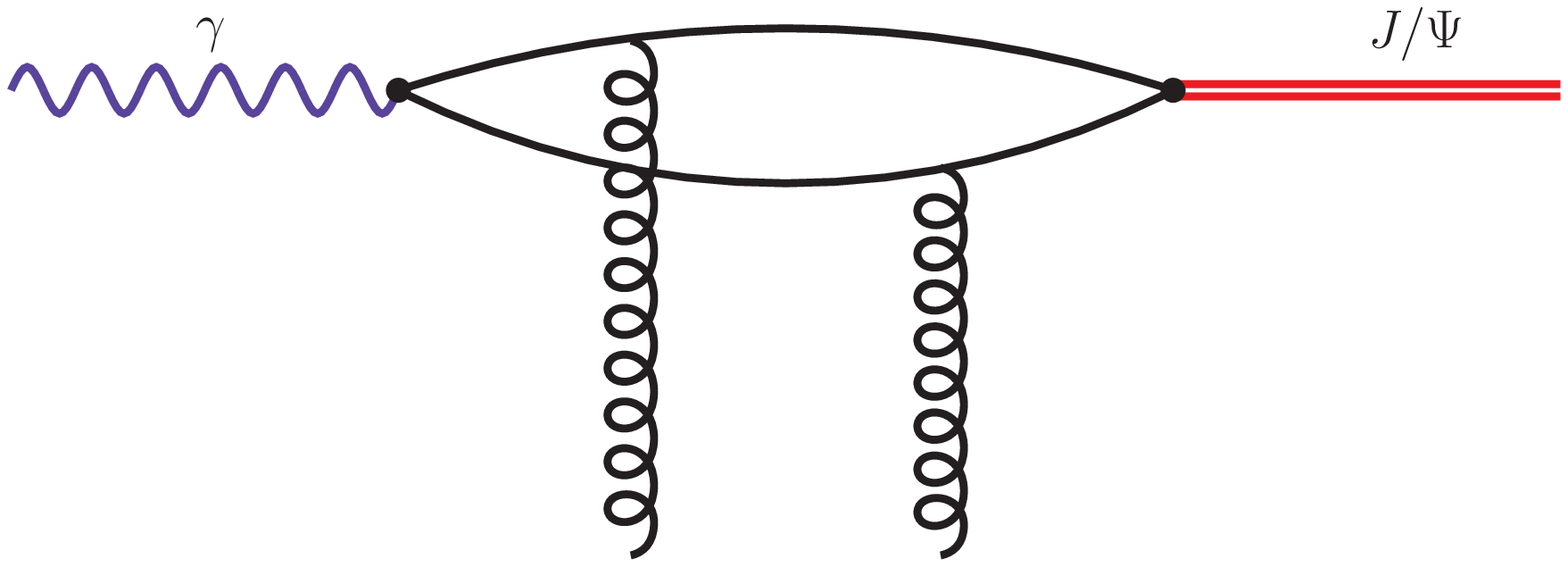}
\includegraphics[width=5cm]{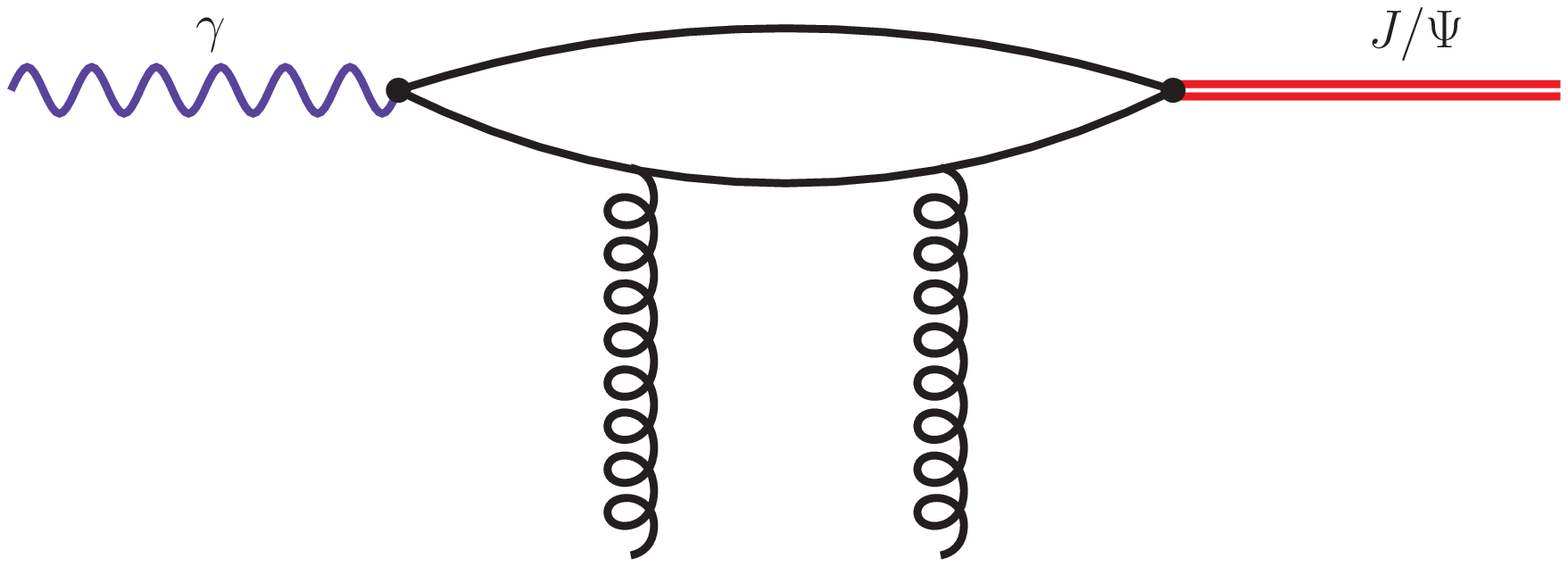}
\end{center}
   \caption{
\small Graphical representation of the two-gluon impact factor for 
the $\gamma \to J/\psi$ transition.
}
 \label{fig:two-gluon_impact_factor}
\end{figure}

The altogether 16 diagrams of the type shown in 
Fig.\ref{fig:two-gluon_exchange} can be 
lead to the amplitude, which can be cast into the impact-factor 
representation:
\begin{eqnarray}
A(\gamma_{\lambda_1} \gamma_{\lambda_2} 
\to V_{\lambda_3} V_{\lambda_4}; s,t) = i s 
\int d^2 \bkappa \, 
{ \Jot( \gamma_{\lambda_1} \to V_{\lambda_3}; \bkappa,\bq) 
\,  \Jot( \gamma_{\lambda_2} \to V_{\lambda_4}; - \bkappa, - \bq)
\over
[(\bkappa + \bq/2)^2 + \mu_G^2][(\bkappa-\bq/2)^2 + \mu_G^2]
}  \, .
\label{eq:A_2g}
\end{eqnarray}
Here $\bq$ is the transverse momentum transfer, $t \approx - \bq^2$, and
$\mu_G$ is a gluon mass parameter. Notice, that the amplitude is finite at
$\mu_G \to 0$, because the impact factors $\Jot$ vanish for $\bkappa \to \pm \bq/2$.
Graphical representation of the impact factors is shown in 
Fig.\ref{fig:two-gluon_impact_factor}.

The amplitude is normalized such, that 
differential cross section is given by
\begin{eqnarray}
{d \sigma(\gamma \gamma \to VV;s) \over dt} = {1 \over 16 \pi s^2} 
\, {1 \over 4} \sum_{\lambda_i} \Big| 
A(\gamma_{\lambda_1} \gamma_{\lambda_2} 
\to V_{\lambda_3} V_{\lambda_4}; s,t) \Big|^2
\, .
\end{eqnarray}
At small $t$, within the diffraction cone, the cross section
is dominated by the $s$-channel helicity conserving amplitude.
In this case, the explicit form of the impact factor is
\begin{eqnarray}
\Jot( \gamma_{\lambda} \to V_{\tau}; \bkappa,\bq) =
\delta_{\lambda, \tau} \, \sqrt{N_c^2-1} \, \alpha_S \, e_Q 
\sqrt{4 \pi \alpha_{em}} \, \int {dz d^2 \bk \over z (1-z) (2 \pi)^3} 
\, \psi_V(z, \bk) \, I(T,T) \, , 
\end{eqnarray}
where $N_c=3$ is the number of colors, $e_Q$ is the charge of the 
heavy quark, and $\psi_V(z,\bk)$ is the light-cone wave function
of the vector meson. It depends on the light-cone-plus momentum fraction
$z$ carried by the quarks as well as on the quark transverse momentum 
$\bk$. The spinorial structure of the $V \to Q \bar Q$ vertex
has been chosen such that it describes the $S$-wave bound state
of quarks \cite{INS2006}, so that
\begin{eqnarray}
I(T,T) =  m_Q^2 \, \Phi_2 + 
\Big[ z^2 + (1-z)^2 \Big] (\bk\bPhi_1) 
+ {m_Q \over M+ 2 m_Q } 
\Big[ 
\bk^2 \Phi_2 - (2z-1)^2 (\bk\bPhi_1)
 \Big] \, .
\end{eqnarray}
Here $\bPhi_1, \Phi_2$ are shorthand notations for the momentum structures,
corresponding to the four relevant Feynman diagrams:
\begin{eqnarray}
\Phi_2 = -{1 \over (\bl + \bkappa)^2 + m_Q^2}  
-{1 \over (\bl - \bkappa)^2 + m_Q^2} 
+{1 \over (\bl + \bq/2)^2 + m_Q^2}
+{1 \over (\bl - \bq/2)^2 + m_Q^2}
\nonumber \\
\bPhi_1 = -{\bl + \bkappa \over (\bl + \bkappa)^2 + m_Q^2}  
-{\bl - \bkappa \over (\bl - \bkappa)^2 + m_Q^2} 
+{\bl + \bq/2 \over (\bl + \bq/2)^2 + m_Q^2}
+{\bl - \bq/2 \over (\bl - \bq/2)^2 + m_Q^2}
\, ,
\label{eq:impact_factors}
\end{eqnarray}
and we used 
\begin{eqnarray}
\bl = \bk + \Big( z - {1 \over 2} \Big)\bq \, , \,  
M^2 = {\bk^2 + m_Q^2 \over z(1-z)} \, . 
\end{eqnarray}
The electronic decay width $\Gamma(V \to e^+ e^-)$ of the vector
meson constrains the light-cone wave function.
Explicitly, the decay constant $g_V$ is
\begin{eqnarray}
g_V = N_c \,  \int {dz d^2 \bk \over z (1-z) (2 \pi)^3} 
\, \psi_V(z, \bk) \, {2 \over 3} M (M+m_Q) \, ,
\label{eq:gv}
\end{eqnarray}
it is related to the decay width through
\begin{eqnarray}
\Gamma(V \to e^+ e^-) = {4 \pi \alpha_{em}^2 e_Q^2 \over 3 M_V^2} \cdot g_V^2
\, .
\label{eq:width}
\end{eqnarray}

In the present approach we assume real photons and therefore only 
transverse photon and vector-meson polarizations are taken into account.
This is sufficiently good approximation for heavy-ion peripheral
collisions where the nucleus charge form factor selects quasi-real
photons.

The running scale of strong coupling constant for the evaluation of the
two-gluon exchange cross section is taken as:
\begin{eqnarray}
\mu^2 &=& \max\{\kappa^2,\bk^2+m_Q^2\} \; . \nonumber  \\ 
\label{renormalization_scale} 
\end{eqnarray}
For the radial wave function we use a Gaussian parametrization
\begin{eqnarray}
\psi_V(z,\bk) = N \, \cdot \exp[-a p^2] \, , \, 
p^2 ={1 \over 4} \Big(  {\bk^2  + m_Q^2 \over z (1-z) }- 4 m_Q^2 \Big) \, .
\end{eqnarray}
Where the constants $N,a$ are determined from the decay width
(\ref{eq:gv},\ref{eq:width}) and the normalization condition
\begin{eqnarray}
N_c \cdot \int {dz d^2\bk \over (2 \pi)^3 z(1-z)} \, M^2 |\psi_V(z,\bk)|^2 = 1
\, .
\end{eqnarray}
For the case of $J/\psi$ of interest here, we use
$m_Q = m_c = 1.5 \, \rm{GeV}$ and $a$ = 1.37 GeV$^{-2}$. 
In this way we get a good description
of the $\gamma p \to J/\psi p$ experimental data
\cite{Cisek}.

\subsection{Non-relativistic limit}

In previous calculations in the literature the extreme 
non-relativistic approximation appropriate for the weakly bound
state of heavy quarks has been adopted.
In this case any quantum motion in the bound state is neglected,
and one uses effectively:
\begin{eqnarray}
\psi_V(z,\bk) = C \cdot \delta(z - {1 \over 2}) \, 
\delta^{(2)}(\bk) \, .
\end{eqnarray}
Notice that this means that quark and antiquark momenta 
in the boosted bound state are collinear.
The constant $C$ results as
\begin{eqnarray}
C = {2 \pi^3 \over N_c M_V^2} \, \sqrt{3M_V \over \pi} \cdot {\cal{R}}(0) = 
{1 \over e_Q \alpha_{em}} {\pi^3 \over N_c M_V} 
\sqrt{3 M_V \Gamma(V \to e^+ e^-) \over \pi}
\, .
\end{eqnarray}
Furthermore, for the weakly bound state $M_V = 2 m_Q$. 
In the impact factor, we can now approximate $I(T,T) = m_Q^2 \Phi_2$, and the
structure $\Phi_2$ simplifies to (hereafter NR means non-relativistic):
\begin{eqnarray}
\Phi^{\mathrm{NR}}_2 =2 \cdot\Big\{
{1 \over \bkappa^2 + M_V^2/4}  - {4 \over \bq^2 + M_V^2} 
\Big\} 
\, .
\end{eqnarray}
The impact factor thus becomes:
\begin{eqnarray}
\Jot^{\mathrm{NR}}( \gamma_{\lambda} \to V_{\tau}; \bkappa,\bq) =
\delta_{\lambda, \tau} \, {\sqrt{N_c^2-1} \over N_c} \, \alpha_S
\, {M_V \over 4}  \, \sqrt{{3 M_V \Gamma(V \to e^+ e^-) \over \alpha_{em}}} 
\cdot 
\Phi^{\mathrm{NR}}_2 \, . 
\end{eqnarray}

\subsection{Meson exchange processes}

The $J/\psi$ meson decays into a photon and pseudoscalar meson
\cite{PDG}. The corresponding
partial decay width can be used to calculate corresponding coupling constant
$g_{\gamma M J/\psi}^2 \propto \Gamma_{J/\psi \to \gamma M(0^-)}/p_{dec}^3$.

The amplitude for the $\gamma \gamma \to J/\psi J/\psi$ due to
pseudoscalar meson ($M$) exchange shown in Fig.\ref{fig:meson_exchange} 
can be written as:
\begin{eqnarray}
{\cal M}_{\lambda_1 \lambda_2 \to \lambda_3 \lambda_4}^{M}(k_1,k_2,p_1,p_2)
&=& g_{\gamma M J/\psi} \epsilon^{\mu_1 \nu_1 \alpha_1 \beta_1}
    \epsilon_{\mu_1}(\lambda_1) \epsilon_{\nu_1}^*(\lambda_3) 
    k_{1,\alpha_1} p_{1,\beta_1} \nonumber \\
&&  F_{\gamma M J/\psi}(t) \frac{-i}{t - m_M^2} F_{\gamma M J/\psi}(t)
    \nonumber \\  
&& g_{\gamma M J/\psi} \epsilon^{\mu_2 \nu_2 \alpha_2 \beta_2}
   \epsilon_{\mu_2}(\lambda_2) \epsilon_{\nu_2}^*(\lambda_4)
   k_{2,\alpha_2} p_{2,\beta_2} \nonumber \\
&+& g_{\gamma M J/\psi} \epsilon^{\mu_1 \nu_1 \alpha_1 \beta_1}
    \epsilon_{\mu_1}(\lambda_1) \epsilon_{\nu_1}^*(\lambda_4) 
    k_{1,\alpha_1} p_{2,\beta_1} \nonumber \\
&&  F_{\gamma M J/\psi}(u) \frac{-i}{u - m_M^2} F_{\gamma M J/\psi}(u)
    \nonumber \\  
&& g_{\gamma M J/\psi} \epsilon^{\mu_2 \nu_2 \alpha_2 \beta_2}
   \epsilon_{\mu_2}(\lambda_2) \epsilon_{\nu_2}^*(\lambda_3)
   k_{2,\alpha_2} p_{1,\beta_2} \; .
\label{amplitude_meson_exchange}
\end{eqnarray}
Above we add $t$ and $u$ meson exchange amplitudes. 
Compared to the $J/\psi \to \gamma M$ decay in the 
$\gamma \gamma \to J/\psi J/\psi$ process the exchanged mesons 
($M = \pi, \eta, \eta_c$) are off mass shell.
This requires to introduce corresponding form factors ($F$ in 
(\ref{amplitude_meson_exchange}) ).
In the following they are parametrized in the exponential form as:
\begin{equation}
F(k^2) = \exp\left( \frac{k^2 - m_{M}^2}{2 \Lambda^2} \right) \; ,
\label{meson_formfactors}
\end{equation}
where $k$ is four-momentum of the exchanged meson.
We take $\Lambda$ = 1 GeV to estimate the meson exchange contributions.
We have found that the cross section for the exchange of $\pi^0$,
$\eta$ and $\eta_c$ is very small, several orders of magnitude
smaller than the cross section for the box and two-gluon exchange
mechanisms. This is very different than for the production of pairs of light
mesons in photon-photon scattering where the exchange of $t$ or $u$
channel mesons lead to relatively big cross section of the order of 1-10 nb.

The contributions of $\pi$ and $\eta$ meson exchange are small because
corresponding coupling constants $J/\psi$-meson-photon are small
compared to the case of $\rho$ or $\omega$ decays as can
be deduced from the partial decay widths. The contribution of
$\eta_c$ exchange is suppressed by large $\eta_c$ mass in the propagator. 

\subsection{Equivalent Photon Approximation
for $A A \to A A J/\psi J/\psi$}

The total cross section can be calculated by the convolution:

\begin{eqnarray}
\sigma\left(AA \rightarrow J/\psi J/\psi AA ;s_{AA}\right) = 
\int dx_1 dx_2  \sigma \left(\gamma\gamma\rightarrow J/\psi J/\psi;
 x_1 x_2 s_{AA} \right) \, 
{dn_{\gamma\gamma}\left(x_1,x_2,{\bf b}\right) \over dx_1 dx_2} .
\end{eqnarray}

The effective Weizs\"acker--Williams photon fluxes can be calculated from:
\begin{eqnarray}
dn_{\gamma \gamma} (x_1,x_2,\bfb) = \int d^2\bfb_1 d^2\bfb_2 \, 
\delta^{(2)}(\bfb - \bfb_1 + \bfb_2) S^2_{\rm{abs}}(\bfb) \,   
{1 \over \pi} {d x_1 \over x_1} 
|\bE(x_1,\bfb_1)|^2 \, {1 \over \pi} {d x_2 \over x_2} 
|\bE(x_2,\bfb_2)|^2 \nonumber \\
\end{eqnarray}
The presence of the absorption factor $S^2_{\rm{abs}}\left({\bf b} \right)$ 
assures that we consider only peripheral collisions, when the nuclei 
do not undergo nuclear breakup. Following \cite{Baur_Ferreira} this can be
taken into account as:  
\begin{eqnarray}
 S^2_{\rm{abs}}\left({\bf b} \right)=
\theta \left({\bfb}-2R_A \right) = \theta \left(|{\bfb}_1-{\bfb}_2|-2R_A \right) \; .
\end{eqnarray}
Thus in the present case, we concentrate on processes with final 
nuclei in the ground state. 
The electric fields can be expressed through the charge 
form factor of the nucleus:
\begin{eqnarray}
 {\bE}\left(x,{\bfb}\right) = Z \sqrt{4\pi \alpha_{em}} 
\int \frac{d^2 {\bq}} {\left(2\pi^2\right)} 
e^{-i \bq \bfb}\frac{{\bq}}{{\bq}^2+x^2M_A^2}
F_{em}\left({\bq}^2+x^2M_A^2\right).
\end{eqnarray}

Different forms of charge form factors are used in the literature. 
Here we use charge form factor calculated as a Fourier transform of
the charge distribution. It was shown in Ref.\cite{KSS2009,KS2010} to be
crucial to take the realistic form factors.

\section{Results}

\subsection{$\gamma \gamma \to J/\psi J/\psi$ reaction}

\begin{figure}[!h]
\includegraphics[width=6cm]{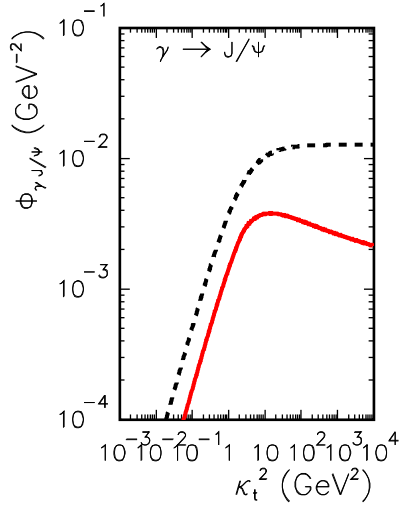}
   \caption{
\small Forward impact factor for $\gamma \to J/\psi$ as a function of
gluon momentum transfer squared. The solid line is our new result 
within the relativistic approach. 
Here $\Phi_{J/\psi}(\kappa^2) \delta_{\lambda_1 \lambda_3} = \Jot( \gamma_{\lambda_1} \to V_{\lambda_3}; \bkappa,\bq=0)$
(see Eq.(\ref{eq:A_2g})). For comparison we also show the 
impact
factor in the non-relativistic approach (dashed line).
}
 \label{fig:forward_impact_factor}
\end{figure}

Let us start from presenting our relativistic impact factors. In 
Fig.\ref{fig:forward_impact_factor} we show the forward ($\bq^2$ =
0) impact factor as a function of gluon transverse momentum squared. 

One can observe a quick rise of the impact factor till gluon momentum transfer
$\kappa^2 \approx$ 3 GeV$^2$. The impact factor obtained in the present
relativistic approach (solid line) is much smaller than that obtained in the
non-relativistic approach (dashed line) used in the literature. This will
lead to corresponding differences in the amplitude of 
the $\gamma \gamma \to J/\psi J/\psi$ process and related cross section.

\begin{figure}[!h]
\includegraphics[width=6cm]{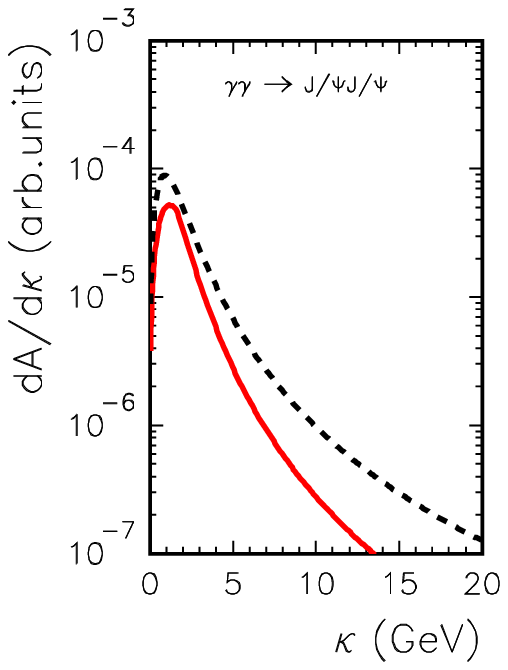}
\includegraphics[width=6cm]{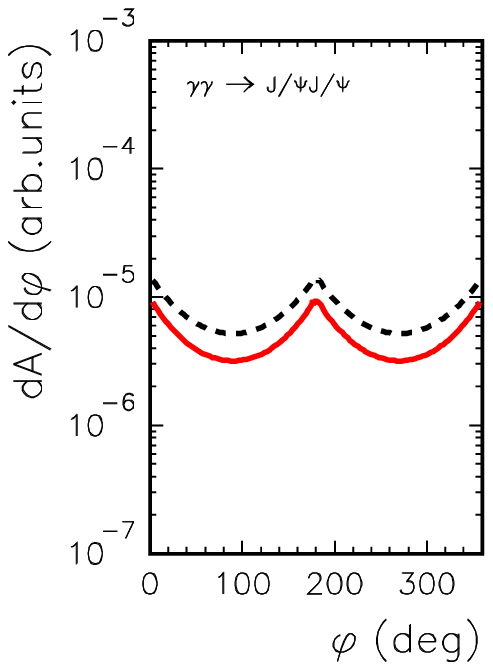}
   \caption{
\small Integrands of the amplitude as a function of $\kappa$ for
$q_t^2$ = 0 (left panel) and as a function of $\phi$ for $q_t^2$ = 5 GeV
(right panel). The solid line is our new result 
within the relativistic approach. Form comparison we show results
for the non-relativistic approach (dashed line).
}
 \label{fig:integrands_of_amplitude}
\end{figure}

Let us illustrate the situation for the amplitude. In 
Fig.\ref{fig:integrands_of_amplitude} we show rather
integrands: $d A / d\kappa$ (left panel) and $d A / d \phi$ (right panel)
instead of the whole amplitude. Here $\phi$ is azimuthal angle between
two-dimensional vectors $\bkappa$ and $\bq$ .
This figure demonstrates also
that it is relatively easy to get numerical convergence
of the integration in calculating the amplitude of the process, 
the decrease in $\kappa$ is rather fast, the
oscillation in $\phi$ is relatively shallow.

\begin{figure}[!h]
\includegraphics[width=6cm]{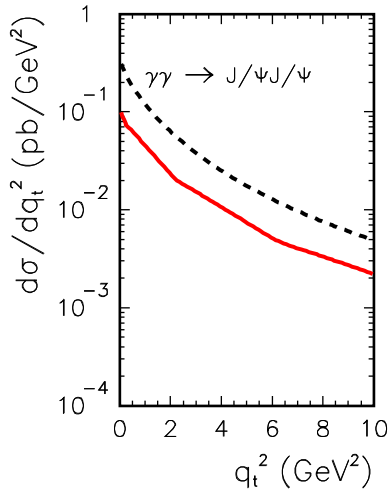}
   \caption{
\small Dependence of the $\gamma \gamma \to J/\psi J/\psi$ cross section
on momentum transfer squared. The solid line represents relativistic
approach and the dashed line the non-relativistic approach.
}
 \label{fig:psig_dqt2}
\end{figure}

In Fig.\ref{fig:psig_dqt2} we show the dependence of the cross section
on two-momentum transfer squared at energy $\sqrt{s} \gg 2 m_{J/\psi}$. 
These two-gluon exchange distributions scale at larger energies. 
At smaller energies kinematical limitations due to
energy-momentum conservation as well as interference of $t$ and $u$
diagrams due to identity of particles (two identical $J/\psi$'s) have 
to be included which destroys the scaling behavior.
Fortunately both effects happen when $W <$ 10 GeV, i.e. when the
contribution of the two-gluon exchange is much smaller than that for the
box mechanism as will be shown below.
The power-like distribution shown in the figure strongly deviates from 
exponential dependence which was assumed for simplicity in
previous calculations (see e.g. \cite{GS2005,GM2007}).

A brief discussion of the infrared sensitivity of the cross section is in
order. Certainly, the cross section is infrared safe -- the impact factors
vanish, when the transverse momenta of $t$-channel gluons go to zero. 
However this does not mean that there is no numerical sensitivity to the
domain of small gluon transverse momenta, where the use of perturbation
theory is not warranted. To investigate this sensitivity, we introduced 
a mass for the $t$-channel gluons. Notice that this mass does not affect
the gauge invariance driven cancellation of impact factors which guarantee
finiteness of the cross section.

Numerically, the infrared sensitivity is quite strong:
\begin{eqnarray}
{\sigma(\gamma \gamma \to J/\psi J/\psi; \mu_G = 0.7 \, {\rm{GeV}}) \over 
\sigma(\gamma \gamma \to J/\psi J/\psi; \mu_G = 0 \, {\rm{GeV}})} \approx 0.45
\, .
\end{eqnarray}

Here, the value $\mu_G = 0.7 \, \rm{GeV}$ corresponds to a vacuum correlation
length of gluons of $R_c \sim 0.29 \, \rm{fm}$. It is motivated by a number
of phenomenological approaches to high-energy scattering 
\cite{Nikolaev:1993ke,Kopeliovich:2007pq}. An analysis of gluon field-strength
correlators on the lattice \cite{D'Elia:1997ne} also suggests a correlation
length in this ballpark. 
Our observation of a surprisingly strong infrared sensitivity of the 
$\gamma \gamma \to J/\psi J/\psi$ cross section agrees with Ref.\cite{GayDucati:2001ud},
which uses the non-relativistic limit of the amplitude. 

\begin{figure}[!h]
\includegraphics[width=.5\textwidth]{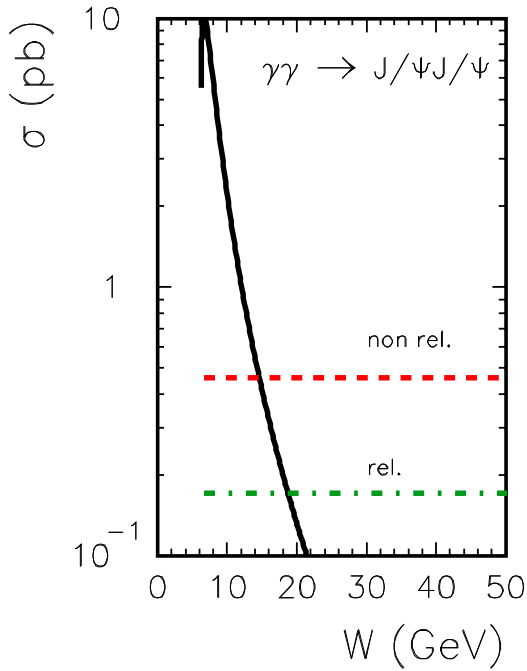}
   \caption{
Energy dependence of the $\gamma \gamma \to J/\psi J/\psi$ cross
section. The solid line is for the the box contribution (dashed line)
while the dashed and dash-dotted lines for two-gluon exchange
contribution in non-relativistic and relativistic approaches, respectively.
}
 \label{fig:dsigma_W}
\end{figure}

The total cross section for the $\gamma \gamma \to J/\psi J/\psi$
reaction is shown in Fig.\ref{fig:dsigma_W}. We show separate
contributions of the box diagrams (dashed line) as well as of
the two-gluon exchange (dotted line). The second contribution scales
at larger energy. Any higher-order interaction (t-channel ladder
exchange) would increase the Born cross section. Unfortunately the existing
calculation in the literature are limited to leading-order BFKL type 
of calculation \cite{KM98,GS2005} which is 
known at present to strongly overestimate the cross section
for other reactions.
We leave a realistic evaluation of the increase of the cross section 
for future studies.

The box and two-gluon exchange cross sections have quite different
energy dependence. As a consequence they are of similar size only in a
very limited range of energy. Therefore in practice the interference
effect can be neglected.

The suppression of the relativistic amplitude in comparison to its 
non-relativistic limit can be traced back to the suppression from
the off-shell quark propagators in Eq.(\ref{eq:impact_factors}).

\subsection{$AA \to A J/\psi J/\psi A$ reaction}

\begin{figure}[!h]
\includegraphics[width=.5 \textwidth]{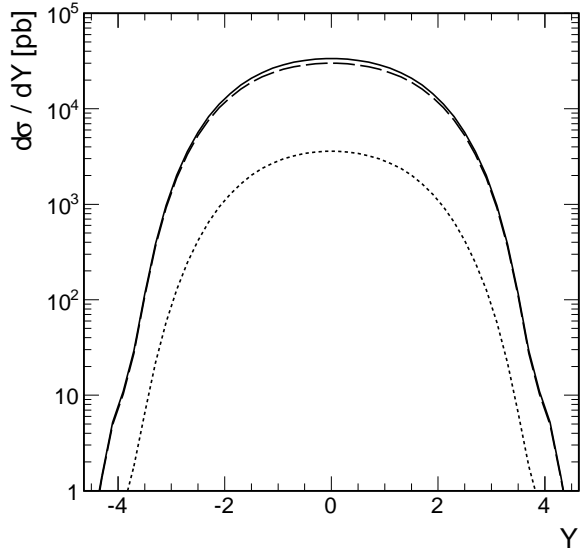}
   \caption{
\small Rapidity distribution of the $J/\psi J/\psi$ pairs in
$^{208}Pb+^{208}Pb \to ^{208}Pb + J/\psi J/\psi + ^{208}Pb$ process for 
$W_{NN}$ = 2.76 TeV. The dashed line is for the box contribution while
the dotted line corresponds to the two-gluon exchange.
}
 \label{fig:dsigma_dYpair}
\end{figure}

\begin{figure}[!h]
\includegraphics[width=.5 \textwidth]{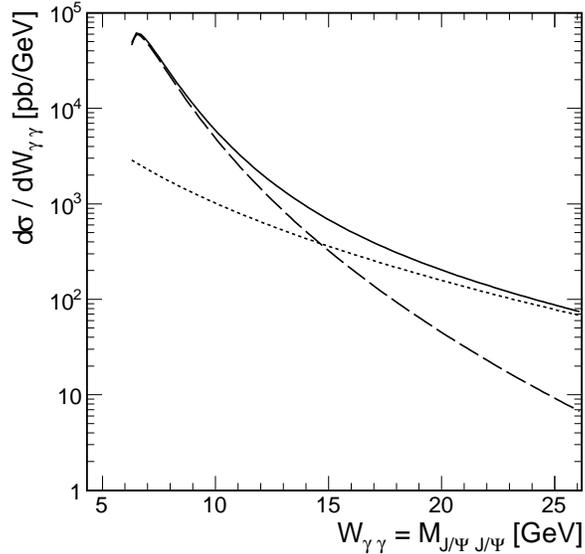}
   \caption{
\small Invariant mass $J/\psi J/\psi$ pair distribution in
$^{208}Pb+^{208}Pb \to ^{208}Pb + J/\psi J/\psi + ^{208}Pb$ process for
$W_{NN}$ = 2.76 TeV. The meaning of the curves is the same as in the
previous figure.
}
 \label{fig:dsigma_dMpair}
\end{figure}

Now we pass to the calculation of the cross section for nuclear
collisions. We follow similar analyses for exclusive production
of $\rho^0 \rho^0$ \cite{KSS2009}, $\mu^+ \mu^-$ \cite{KS2010}, $c \bar c$
\cite{KSMS2011} or $\pi \pi$ \cite{KS2011}. In those approaches a calculation
of the distribution in rapidity of the pair or invariant mass of the
pair is particularly easy. In Fig.\ref{fig:dsigma_dYpair} we show
distribution in rapidity of the pair for $W_{NN}$ = 2.76 TeV. 
Both contributions (box and two-gluon exchange) are shown separately.
The phase
space integrated nuclear cross section is $\sigma$ = 0.1 $\mu$b.
which is of similar size as the cross section for 
production of $D \bar D$ meson pairs \cite{LS2011}.
The contribution of the two-gluon exchange is an order of magnitude
smaller than that for the box mechanisms. No previous evaluation
of the nuclear cross section included the dominant box contribution.
In Fig.\ref{fig:dsigma_dMpair} we show distribution in $J/\psi J/\psi$
invariant mass. This distribution drops quickly as a function of the
two-meson invariant mass, which reflects the fall-off of the 
effective photon-photon luminosity at large $\gamma \gamma$-energies.
The two-gluon exchange contribution shows up only at $M_{J/\psi J/\psi} >$
15 GeV.
The situation can improve slightly if the rise of the $\gamma \gamma$
cross section due to BFKL-type effects were included. 
We do not expect, however, that a realistic
calculation including a BFKL-type ladder interaction between two 
$c \bar c$ pairs would change our predictions significantly for invariant
$J/\psi J/\psi$ masses smaller than say 50 GeV.

\begin{figure}[!h]
\includegraphics[width=.5 \textwidth]{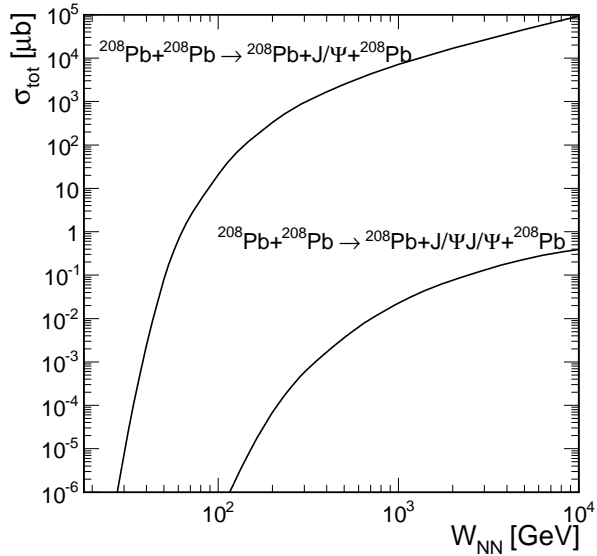}
   \caption{
\small Energy dependence of the cross section for 
the $^{208}Pb+^{208}Pb \to ^{208}Pb + J/\psi J/\psi + ^{208}Pb$ process
versus that for the $^{208}Pb+^{208}Pb \to ^{208}Pb + J/\psi + ^{208}Pb$
process. The latter cross section is taken from \cite{CSS2012}.
}
 \label{fig:dsigma_WNN}
\end{figure}

Recently we have calculated the cross section for exclusive production
of single $J/\psi$ meson \cite{CSS2012}. In Fig.\ref{fig:dsigma_WNN}
we compare the cross section for the pair production (photon-photon
fusion) with that for the production of a single $J/\psi$ via the 
``photon-pomeron'' mechanism.
The latter cross section is more than five orders of magnitude larger.
The single $J/\psi$ production was measured e.g. at RHIC \cite{RHIC_Jpsi}.
Similar analysis is being performed by the ALICE collaboration at the LHC
\cite{ALICE_Jpsi}. A present statistics is a few hundreds of $J/\psi$
\cite{Mayer}.
This means that with present statistics the number of measured pairs
would be less than one.
Clearly a better statistics is necessary to see pairs of $J/\psi$.
A better situation could be for the $p p \to p p J/\psi J/\psi$ reaction
where the effective $\gamma \gamma$ luminosity at large $J/\psi J/\psi$ 
invariant masses is much larger.

If the nuclear cross section for 
$^{208}Pb + ^{208}Pb \to ^{208}Pb (J/\psi J/\psi) ^{208}Pb$
was much larger than predicted in this paper it could mean that a double
scattering photon-pomeron mechanism discussed in Ref.\cite{CSS2012} plays
important role. A measurement of pair production at ALICE can therefore
provide a quite new information. The exclusive nuclear double-scattering
production of mesons was not discussed so far in the literature.

\section{Conclusions}

In the present paper we have calculated the total cross section and
angular distributions for the $\gamma \gamma \to J/\psi J/\psi$ process.
For the first time we have included both formally two-loop contribution,
called here ``box contribution'', included so far in only one previous study,
as well as two-gluon exchange contribution, formally higher order, 
discussed already in the literature.
We have clarified a disagreement of results of different calculations in 
the literature. In our approach the two-gluon exchange contribution has
been calculated in a relativistic approach with the wave function
adjusted to reproduce the HERA data for $\gamma p \to J/\psi p$ reaction.
For the first time we have calculated relativistic off-forward two-gluon
exchange amplitude. We have quantified a difference between the relativistic
and non-relativistic approach used so far in the literature.
The relativistic approach with ``realistic'' quark-antiquark $J/\psi$
wave function gives smaller cross sections than those obtained in
non-relativistic approximation.
In addition, we have also estimated a contribution of pseudoscalar 
($\pi^0, \eta, \eta_c$) meson exchange mechanism. The latter ones turned
out to be practically negligible.

The elementary cross section has been used to calculate cross section
for ultra-peripheral ultra-relativistic heavy ion collisions
($A A \to A J/\psi J/\psi A$) at the LHC. In this context we have used
equivalent photon approximation in the impact parameter representation
and realistic charge form factor of the $^{208} Pb$ nucleus.
The nuclear cross section has been estimated and distribution in
the $J/\psi J/\psi$ rapidity and invariant mass have been presented. 
The cross section for double $J/\psi$ production was compared with
that for single $J/\psi$ production from \cite{CSS2012}. At the LHC
energies the double $J/\psi$ production cross section is smaller by more
than five orders of magnitude compared to that for single $J/\psi$ 
production. 
A measurement of the $J/\psi J/\psi$ pair production may be then rather 
difficult as one has to take into account in addition a product of respective
branching fractions for decays into lepton-antilepton pairs.
On the other hand much bigger cross section could be a signal of
double scattering ``sequential'' production of two single $J/\psi$'s.

In the present analysis we have concentrated on the photon-photon
mechanism of the $J/\psi J/\psi$ pair production. Evidently similar 
mechanisms exist for gluon-gluon (sub)processes, leading to the inclusive
production of $J/\psi J/\psi$ pairs in proton-proton collisions. 
Such a production has been observed recently by the LHCb collaboration 
\cite{LHCb}.
Only the box-type diagrams were included in corresponding calculations 
\cite{KKS2011,BSZ2011}. The two-gluon exchange mechanism considered
here may in principle compete with the double-parton production
mechanism of the $J/\psi J/\psi$ pairs discussed recently in 
\cite{KKS2011,BSZ2011}.

After we have performed the calculation presented here we have found a
new preprint \cite{Ahmadov:2012dn} where the production of two vector mesons is
discussed. Some additional photon-exchange processes are considered. In the
$\gamma \gamma$ channel, the authors do not discuss the dominant box-mechanism.
Also, nuclear charge form factors which are important at large $\gamma \gamma$
energies \cite{KSS2009,KS2010} are neglected.

\vspace{1cm}

{\bf Acknowledgments}

We are indebted to Christoph Mayer for a discussion of the ALICE
measurement of the $J/\psi$ meson.
This paper was partially supported by the Polish
Grants DEC-2011/01/B/ST2/04535 and N N202 236 640.



\begin{thebibliography}{100}

\bibitem{GPS88}
  I.~F.~Ginzburg, S.~L.~Panfil and V.~G.~Serbo,
  Nucl.\ Phys.\  B {\bf 296}, 569 (1988).

\bibitem{KM98}
J.~Kwieci\'nski and L. Motyka, Phys. Lett. {\bf B438}, 203 (1998).


\bibitem{GS2005}
  V.~P.~Goncalves and W.~K.~Sauter,
  Eur.\ Phys.\ J.\  C {\bf 44}, 515 (2005)
  .

\bibitem{GM2007}
  V.~P.~Goncalves and M.~V.~T.~Machado,
  Eur.\ Phys.\ J.\  C {\bf 49}, 675 (2007).

\bibitem{Qiao}  
 C.-F.\ Qiao, \prd{64}{077503}{2001} .

\bibitem{NRQCD}
 C.~-H.~Chang,
  Nucl.\ Phys.\ B {\bf 172}, 425 (1980);
R.~Baier and R.~R\"uckl,
Phys.\ Lett.\ B {\bf 102}, 364 (1981);
  E.~L.~Berger and D.~L.~Jones,
  Phys.\ Rev.\ D {\bf 23}, 1521 (1981).


\bibitem{PDG}
J. Beringer et al. (Particle Data Group), Phys.\ Rev.\ D {\bf 86}, 010001 (2012).


\bibitem{FORM} J.\ A.\ M.\ Vermaseren,
       {\it Symbolic Manipulations with FORM}
       (Computer Algebra Nederland, Kruislaan, SJ Amsterdaam, 1991,
        ISBN 90-74116-01-9).

\bibitem{INS2006}
  I.~P.~Ivanov, N.~N.~Nikolaev and A.~A.~Savin,
  Phys.\ Part.\ Nucl.\  {\bf 37}, 1 (2006).

\bibitem{Cisek} 
 A.~Cisek, \textit{Exclusive processes with large rapidity gaps in the 
formalism of unintegrated gluon distributions}, (PhD thesis, The Henryk Niewodnicza\'nski
Institute of Nuclear Physics, Polish Academy of Sciences, Krak\'ow, 2012) 



\bibitem{Baur_Ferreira} 
  G.~Baur and L.~G.~Ferreira Filho,
  Nucl.\ Phys.\ A {\bf 518}, 786 (1990);
R.~N.~Cahn and J.~D.~Jackson,
  Phys.\ Rev.\ D {\bf 42}, 3690 (1990).

\bibitem{KSS2009} 
M. K{\l}usek, W. Sch\"afer and A. Szczurek,
Phys. Lett. {\bf B674} 92, (2009). 

\bibitem{KS2010}
M. K{\l}usek-Gawenda and A. Szczurek,
Phys. Rev. {\bf C82} 014904, (2010).

\bibitem{Nikolaev:1993ke} 
  N.~N.~Nikolaev, B.~G.~Zakharov and V.~R.~Zoller,
  JETP Lett.\  {\bf 59}, 6 (1994);
N.~N.~Nikolaev, B.~G.~Zakharov and V.~R.~Zoller,
  JETP Lett.\  {\bf 66}, 138 (1997)
  [Pisma Zh.\ Eksp.\ Teor.\ Fiz.\  {\bf 66}, 134 (1997)];
  N.~N.~Nikolaev, W.~Sch\"afer, B.~G.~Zakharov and V.~R.~Zoller,
  JETP Lett.\  {\bf 84}, 537 (2007).

\bibitem{Kopeliovich:2007pq} 
  B.~Z.~Kopeliovich, I.~K.~Potashnikova, B.~Povh and I.~Schmidt,
  Phys.\ Rev.\ D {\bf 76}, 094020 (2007).

\bibitem{D'Elia:1997ne} 
  M.~D'Elia, A.~Di Giacomo and E.~Meggiolaro,
  Phys.\ Lett.\ B {\bf 408}, 315 (1997).

\bibitem{GayDucati:2001ud} 
  M.~B.~Gay Ducati and W.~K.~Sauter,
  Phys.\ Lett.\ B {\bf 521}, 259 (2001).

\bibitem{KSMS2011}
M. K{\l}usek-Gawenda, A. Szczurek, M. Machado and V. Serbo,
Phys. Rev. {\bf C83} 024903, (2011).

\bibitem{KS2011} 
M. K{\l}usek-Gawenda and A. Szczurek,
Phys. Lett. {\bf B700} 322, (2011).

\bibitem{LS2011} 
M. {\L}uszczak and A. Szczurek,
Phys. Lett.{\bf B700} 116, (2011).

\bibitem{CSS2012}
A. Cisek, W. Sch\"afer and A. Szczurek, 
Phys.\ Rev.\ C {\bf 86}, 014905 (2012).

\bibitem{RHIC_Jpsi}
S. Afanasiev et al. (PHENIX Collaboration), Phys. Lett. {\bf B679} 321,
(2009).

\bibitem{ALICE_Jpsi}
A. Nystrand, a talk at the Blois workshop, Qui Nhon, Vietnam, December 2011.

\bibitem{Mayer}
Ch.~Mayer, \textit{private communication}, July 2012.

\bibitem{LHCb}
R. Aaij et al. (LHCb collaboration), Phys. Lett. {\bf B707} 52, (2012).

\bibitem{KKS2011}
C.H. Kom, A. Kulesza and W.J. Stirling, Phys. Rev. Lett. {\bf 107}
082002, (2011).

\bibitem{BSZ2011}
S.P. Baranov, A.M. Snigirev and N.P. Zotov, Phys. Lett. {\bf B705}
116, (2011).


\bibitem{Ahmadov:2012dn} 
  A.~I.~Ahmadov, B.~A.~Kniehl, E.~A.~Kuraev and E.~S.~Scherbakova,
  arXiv:1206.4441 [hep-ph].

\end{thebibliography}
\end{document}